\documentclass[aps,pre,epsf,superscriptaddress,amsmath,amssymb,amsfonts,twocolumn,showpacs]
{revtex4-1}
\usepackage{adjustbox}
\usepackage{tikz}
\usepackage{xcolor}
\usepackage{bbm}

\usepackage{amsmath,amssymb,amsfonts,bm,comment}
\usepackage{graphics,float,tikz}
\usepackage{appendix}
\usetikzlibrary{arrows.meta,arrows}
\usepackage[caption=false]{subfig}
\usepackage[colorlinks=true,
	linkcolor=blue,
	filecolor=magenta,      
	urlcolor=blue,
	citecolor=blue]{hyperref}
\usepackage{physics,xparse,mathtools,isotope}
\usepackage[table-parse-only]{siunitx}
\usepackage{multirow,array,relsize,lipsum,longtable,tabularx}
\usepackage{dsfont}

\usepackage{amsthm}



\newcommand{\gniewko}[1]{\color{black}{#1}\color{black}}
\newcommand{\madhura}[1]{\color{black}{#1}\color{black}}
\newcommand{\rethink}[1]{\color{black}{#1}\color{black}}

\begin{document}
\title{Signatures of superradiance in intensity correlation measurements in a two-emitter solid-state system}
	\author{Madhura Ghosh Dastidar}
 
    \affiliation{Quantum Center of Excellence for Diamond and Emerging Materials (QuCenDiEM) Group, Department of Physics, Indian Institute of Technology Madras, Chennai 600036, India}

	\author{Aprameyan Desikan}

	\affiliation{Department of Physical Sciences, Indian Institute of Science Education \& Research Mohali Sector 81 SAS Nagar, Manauli PO 140306 Punjab, India}

	\author{Gniewomir Sarbicki}
 
	\affiliation{Institute of Physics, Faculty of Physics, Astronomy and Informatics, Nicolaus Copernicus University, Grudzi\c adzka 5/7, 87-100 Toru\'n, Poland}

	\author{Vidya Praveen Bhallamudi}
\email[]{praveen.bhallamudi@iitm.ac.in}
\affiliation{Quantum Center of Excellence for Diamond and Emerging Materials (QuCenDiEM) Group, Department of Physics, Indian Institute of Technology Madras, Chennai 600036, India}
 
	\date{\today}
 
	\begin{abstract}
We perform intensity correlation ($g^{(2)}(\tau)$) measurements on nitrogen-vacancy (NV) emitters embedded in diamond nanopillars. We observe an increase in transition rates from both the singlet and triplet states by a factor of $\approx 6$, indicating cooperative effects between the multiple emitters in the pillar, at room temperature. We simultaneously observe a $g^{(2)}(0) > 0.5 (\to 1$) as opposed to $g^{(2)}(0) < 0.5$ for others (and as expected for single emitters), indicating the presence of at least two emitters. Furthermore, we observe a triple exponential behaviour for the $g^{(2)}$ in contrast to the standard double exponential behaviour seen for single NV emitters. To understand our experimental observations, we developed a theoretical model. We solve the Lindblad master equation, tailored for single and two NV centers, to study their dissipative dynamics when coupled to a common electromagnetic field, at a finite temperature. Through this, we identify superradiant emission from a two-emitter system as the most likely explanation for our observed data.  We also find that random number generation using the coupled emitter system performs better under the NIST test suite and explain it in terms of an entropy-driven model for a coupled emitter system. Our results provide a new signature for multiphotonic states, such as superradiant states, using intensity correlation measurements, that will become important for quantum photonic technologies progress. 
\end{abstract}
	
	\maketitle
	\section{Introduction}
 Quantum emitters are crucial to the success of quantum technologies~\cite{wehner2018quantum, lobl2024loss, levonian2022optical}. Single photon emission from various colour centers in diamond and other solid-state materials are well-established~\cite{aharonovich2016solid, he2015single, ren2019review} and are being investigated for quantum networks~\cite{lee2020integrated}.  Certain defects such as nitrogen-vacancy (NV) centers in diamond also offer a spin-photon interface~\cite{atature2018material, castelletto2020hexagonal, shandilya2022diamond}. Going beyond the single photon emission, multiphotonic states need to be studied due to their potential for improving the performance in a quantum system~\cite{paulisch2019quantum, dell2006multiphoton}. 
 
Superradiance is collective emission (of a multiphotonic state) from a system of (indistinguishable) emitters, confined in a volume, with a characteristic dimension comparable to or smaller than the wavelength of excitation. The emitters are coupled to the common electromagnetic field~\cite{dicke1954coherence}. It has multiple applications in quantum memory~\cite{rastogi2022superradiance} and quantum metrology~\cite{paulisch2019quantum}. While, superradiance has been experimentally observed for various systems~\cite{raino2018superfluorescence, cygorek2023signatures, lukin2023two, mlynek2014observation}, it is challenging to observe in solid-state emitters due to the enhanced disspitative mechanisms~\cite{gross1982superradiance, wolters2013measurement}, affecting the indistinguishability of emitters. Further at room temperature, a high density of emitters was required for the emergence of superradiant behaviour~\cite{bradac2017room}. 
 
 A systematic study is challenging in the large number of solid-state emitters' limit; typically, radiative lifetime measurements are used to establish superradiance~\cite{bradac2017room}. A more controlled superradiance from a finite number of emitters is desirable, for understanding such emission in greater detail in solid-state emitters and for reproducible technology. Two-emitter superradiant systems have been experimentally studied~\cite{mlynek2014observation, trebbia2022tailoring}. In particular,~\cite{lukin2023two} shows a bunching effect due to superradiance on the intensity correlation (so-called $g^{(2)}$) function, a canonical measurement for characterising single quantum emitters. However, a detailed theoretical analysis was not presented. Further, only the amplitudes of $g^{(2)}(0)$ was used to characterise the emission.

 Applications of multiphotonic states have been seen in quantum key distribution~\cite{helwig2009multimode,teng2021sending}. However, the randomness of the key is an important aspect of quantum cryptography, thereby causing a need for true generators~\cite{chen2019single}. A primary quantum process and a subsequent measurement provides an inherent randomness to the outcome~\cite{acin2016certified}. The randomness can be quantified using entropic measures or various test suites. 
 

 Here, we report the experimental observation of optical superradiance from two nitrogen-vacancy (NV) centers in a diamond nanopillar and a detailed analytical model based on the dissipative dynamics of such a system. We study the effect of excitation power on the lifetimes involved in the superradiant emission and highlight the qualitative differences from single photon emission. Finally, we discuss the entropy of quantum random number generators using the coupled emitter system and compare it with single emitters.
 
\begin{figure*}[htp]
    \centering    
    \includegraphics[width=0.8\textwidth]{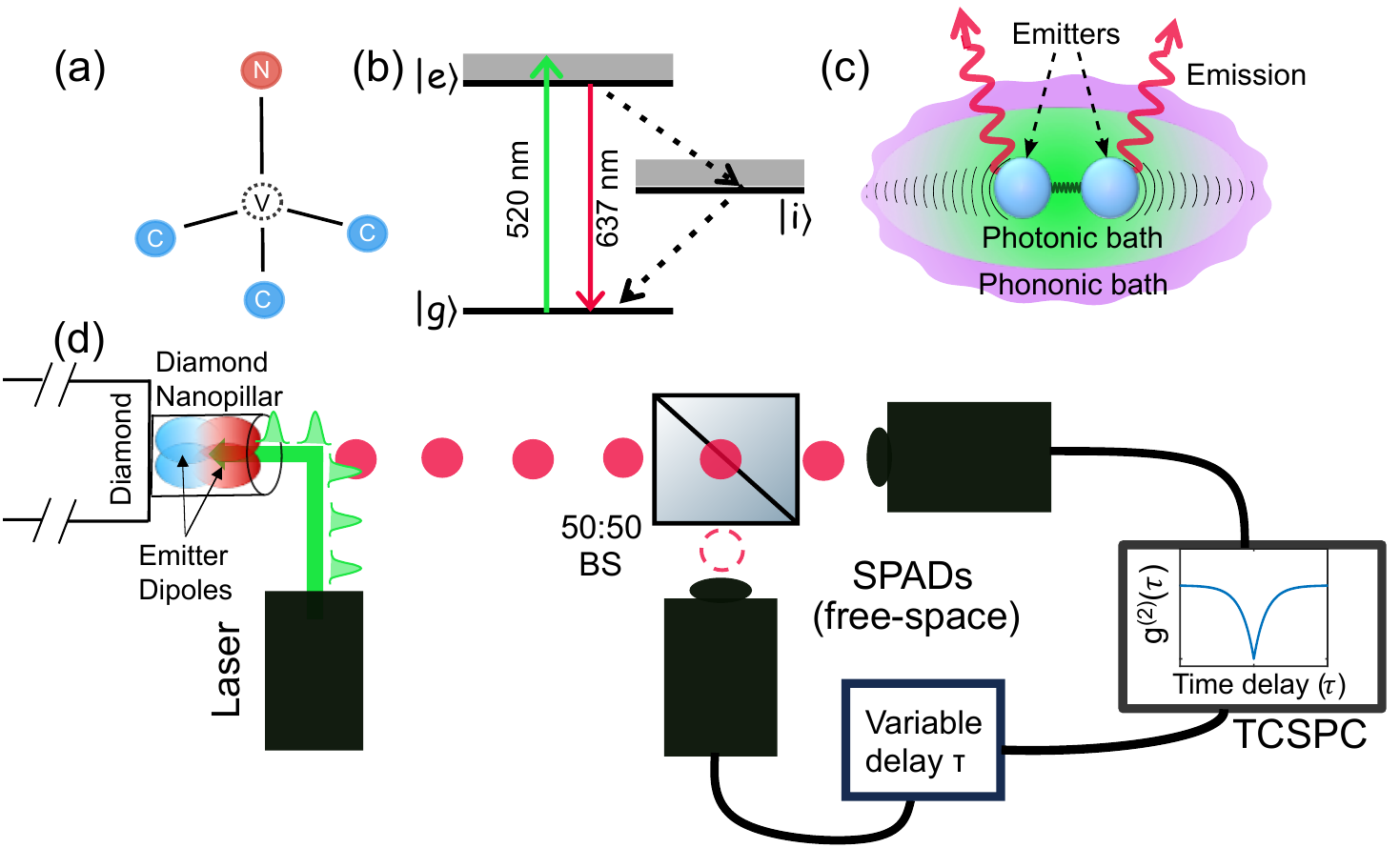}
    \caption{(a) Schematic of the NV center defect in diamond. (b) Three-level energy structure of an NV center. $\ket{g}, \ket{i}, \ket{e}$ denote the ground, intermediate and excited states. the dashed lines indicate non-radiative transitions. The shaded regions indicate the vibronic levels of the excited and intermediate states. (c) Pictorial representation of two coupled NV centers (solid blue circles) in a common photonic bath (electric field in green) and phononic bath (purple). The red squiggly arrows represent the radiative emissions from the emitters. (d) The simplified experimental scheme depicts the Hanbury-Brown Twiss experiment performed for our system. The coupled emitters, embedded in a diamond nanopillar, are perturbed using a laser emitting at 520 nm wavelength and the emission is split using a 50:50 beam-splitter (BS) and sent to two detectors $-$ single-photon avalanche diodes (SPADs). The coincidences (as a function of detection delay on one of the detectors) from the two detectors are measured using a time-correlated single photon counter (TCSPC). An exemplar plot for the normalized coincidences ($g^{(2)}(\tau)$) vs. time delay ($\tau$) for a single emitter is shown as an inset figure on the TCSPC. }
    \label{NV_level_structure}
\end{figure*}

\section{The Nitrogen Vacancy (NV) Center and its Relevant Energy Levels}
We consider the negatively charged, nitrogen vacancy (NV$^-$, which we will refer to as NV) centers in diamond (schematically shown in Fig.~\ref{NV_level_structure} (a)). To ensure the accurate modelling of our system correctly, we derive all possible/accessible energy levels of the NV center. In the tight-binding model, it consists of 6 valence electrons that can be described by 2-hole wavefunctions. First, we derive the one-hole wavefunctions for the system.
The vacancy results in the appearance of four dangling orbitals $-$ three from carbons and one from nitrogen. Each is the lowest-energy solution of a single-electron Schr\"odinger equation with a potential from nuclei and bonded electrons, satisfying the $C_{3v}$ symmetry of the defect. Let us denote these 
3 carbon orbitals and one nitrogen orbital as $\sigma_1, \sigma_2, \sigma_3$ and $\sigma_N$, respectively. 

The single-hole Hamiltonian takes in the $\{\sigma_1, \sigma_2, \sigma_3, \sigma_N\}$ basis a most general form allowed by the symmetry of the defect~\cite{maze2011properties}:

\begin{align}
    V &= V_N\ketbra{\sigma_N} + \sum_i \{V_C \ketbra{\sigma_i} + (h_N \ketbra{\sigma_i}{\sigma_N} +h.c.)\} \nonumber\\
    &+ \sum_{i>j} h_C\ketbra{\sigma_i}{\sigma_j} + h.c. 
    \label{electron_ion_matrix}
\end{align}

On diagonalizing $V$, we obtain the eigenbasis of the single-hole system and restrict considerations to the three lowest levels, denoted further as $\{ \ket{x}, \ket{y}, \ket{1} \}$. Now, using the symmetry group of the system, 
we obtain 9 vectors of the 2-hole eigenbasis from the single-hole eigenstates by applying the projection formula~\cite{weissbluth2012atoms} (see Supplementary Information for details). The resulting energy levels are perturbed by the Coulomb interaction between the 2-hole orbitals. Further, we add the spin degree of freedom, resulting in 15 symmetric 2-hole eigenstates. Finally, we discuss the spin-orbit interactions.


We have comprehensively derived all possible selection rules of radiative and non-radiative transitions between the energy levels by applying the orthogonality theorem from the representation theory to the $C_{3v}$ group (summarised in Fig. S1 of the Supplementary Information). 
\gniewko{The selection rules allow a radiative transition between the ground level and one of the excited ones. A non-radiative transition between them is allowed via one intermediate state, and if the wavelength of the input electromagnetic field is fitted to this radiative transition, the system can be treated as a three-level system spanned by the orbitals } (see Supplementary Information for details):

\begin{align}
        \ket{g} = A_1(3) &= \frac{1}{2}(\ket{xy} - \ket{yx}) \otimes (\ket{\uparrow\downarrow} + \ket{\downarrow\uparrow}) \\
        \ket{i} = A_1(1) &= \frac{1}{2}(\ket{xx} + \ket{yy}) \otimes (\ket{\uparrow\downarrow} - \ket{\downarrow\uparrow}) \\
        \ket{e} = A_1(4) &= \frac 12(
            (\ket{1-}-\ket{-1})\otimes\ket{\uparrow\uparrow}
            \nonumber \\
            & -
            (\ket{1+}-\ket{+1})\otimes\ket{\downarrow\downarrow}
        )
\end{align}

In our experiment, we illuminate a sample containing an NV center with a laser field of wavelength $520$ nm (higher than energy gap between $\ket{e}$ and $\ket{g}$) [see Fig.~\ref{NV_level_structure} (b)].
In this way, the system becomes excited to one of the vibronic levels of $\ket{e}$. Its de-excitation to the ground vibronic level occurs within ps~\cite{ulbricht2018vibrational}, i.e., much shorter than the lifetime of the excited state. Thus, we can neglect the transitions occurring within the vibrational levels.

Next, the system de-excites to the ground state via photon emission, and non-radiatively through the intermediate state $\ket{i}$. The (thermally induced) populations of other states can be neglected because the energy of the laser is much higher than $k_BT$. 

\section{Experimental Observation of Intensity Correlations for Superradiance}

As shown in Fig.~\ref{NV_level_structure} (c), a schematic representation of photonic and phononic dissipation from a two-emitter system occurs when coupled to a common electric field. If the emitters are located in a region smaller than the wavelength of excitation, it can show superradiance.

\begin{figure}[ht]
        \centering
\includegraphics[width=0.4\textwidth]{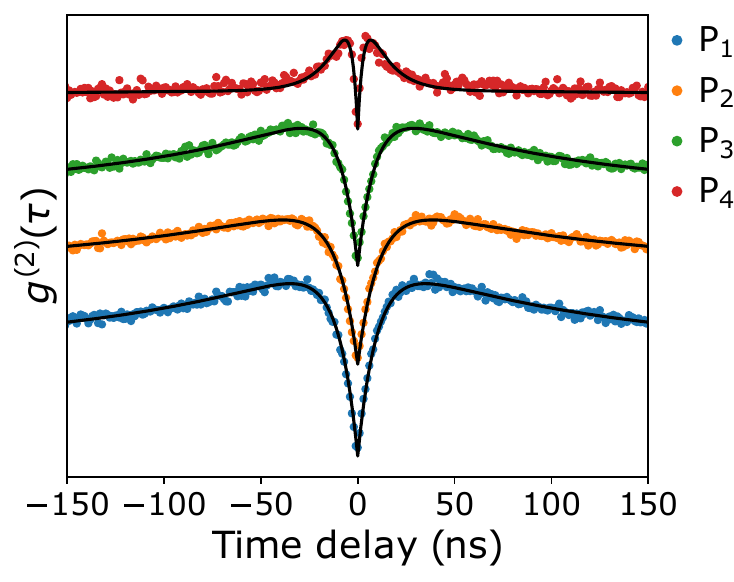}
        \caption{\textbf{Comparison of intensity correlations from single emitters and two emitters: } We show the $g^{(2)}(\tau)$ function (measured: dots; solid line for fitting with Eqs.~\ref{g2_fit_single_emitter} (for $P_1, \dots, P_3$) and~\ref{g2_fit_4level} for $P_4$) for four pillars containing NV centers (labelled as $P_1, P_2, P_3$ and $P_4$). It is plotted as a function of time delay $\tau$ in detecting coincidences. The datasets are stacked vertically by adding a manual offset for better visibility. 
        }
\label{fig:g2Lifetime_allPillars}
    \end{figure}
    

\begin{figure*}[ht]
        \centering
    \includegraphics[width=0.7\textwidth]{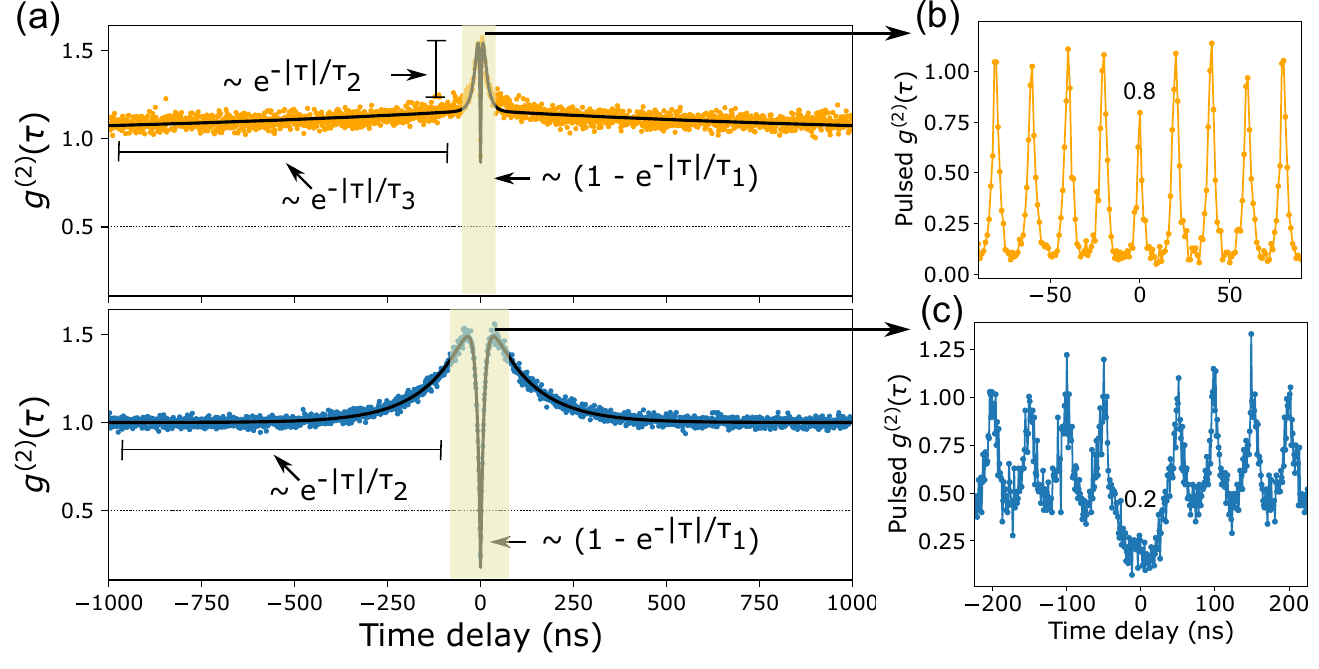}
        \caption{\textbf{Evidence of superradiance: }(a) Expanded view of the cw $g^{(2)}(\tau)$ (measured: dots; fitting: solid line) for $P_4$ and $P_1$ highlighting the regions showing three exponential (top panel) and two exponential (bottom panel) decays, respectively. 
        The dashed line indicates $g^{(2)}(0) = 0.5$. For antibunching $g^{(2)}(0) < 0.5$, which is the case for $P_1$, whereas $g^{(2)}(0) > 0.5$. (b)-(c) Pulsed $g^{(2)}(\tau)$ measurements (pump is pulsed at 50 MHz and 20 MHz, respectively, with $\sim 200$ ps pulses) are performed for the shaded yellow region of the cw $g^{(2)}(\tau)$ from $P_4$ and $P_1$ The functional form of $g^{(2)}(\tau)$ and loss of antibunching $g^{(2)}(0)>0.5$ indicate a coupling of two emitters in $P_4$.}
        \label{fig:g2_P1P4}
\end{figure*}

\subsection{Hanbury-Brown Twiss Experiment: for Intensity Correlation}
To comment on the intensity correlations in the emitted light, we perform the Hanbury-Brown Twiss (HBT) experiment [see Fig.~\ref{NV_level_structure} (d)]. From this experiment, we extract the second-order correlation function $g^{(2)}(\tau)$ w.r.t the delay in detection of coincidences ($\tau$) between two single photon detectors (SPDs) [see Methods Sections]. All the measurements described were conducted multiple times over various days to ensure repeatability. In Fig.~\ref{fig:g2Lifetime_allPillars} (b), we show the $g^{(2)}(\tau)$ vs. time delay $\tau$ for various nanopillars $P_i (i \in [1,4])$ containing emitters. Nanopillars containing single NV centers ($P_1, \dots, P_3$) exhibit the three-level model owing to the intermediate metastable (singlet) state. We thereby use the following fitting function [see Sec.~\ref{One NV Center}]:

\begin{equation}\label{g2_fit_single_emitter}
    g^{(2)}(\tau) = 1-(1+a)e^{-\tau/\tau_1}+ae^{-\tau/\tau_2}
\end{equation}

which we extract from Eq.~\ref{g2_single_emitter} where, $1/\lambda_i = \tau_i, i \in \{1, 2\}$. A clear antibunching dip ($g^{(2)}(0) < 0.5$) was observed for each case of single emitter ($P_1, \dots, P_3$), which is characteristic of single photon emission. Further, we can see two characteristic time constants in the $g^{(2)}(\tau)$, from the two exponential components. The time constants ($\tau_1, \tau_2$) represent the inverse of transition rates of the emitter between different energy levels ($\ket{e} \to \ket{g}, \ket{i} \to \ket{g}$), respectively. For $\tau >> \tau_2$, $g^{(2)}(\tau) \to 1$ asymptotically. We observe that for the 3 single emitters each in nanopillars $P_i, i \in \{1, 2, 3\}$, the three-level equation for $g^{(2)}(\tau)$ fits well to the measured data (see Fig. S2 Supplementary Information). 

Interestingly, for pillar $P_4$, we observe the certain unique properties of the $g^{(2)}(\tau)$ function [see Fig.~\ref{fig:g2_P1P4} (a)]. First, the amplitude of the antibunching dip is close to 0.8, i.e., $g^{(2)}(0) \to 0.8$. Since $g^{(2)}(0) > 0.5$, we consider the dynamics of this system to be similar to that of 2 NV centers. 
While fitting the behaviour of $g^{(2)}(\tau)$ as a function of $\tau$ we use the function:

\begin{equation}\label{g2_fit_4level}
    g^{(2)}(\tau) = 1- (1+a_1+a_2)e^{-\tau/\tau_1} + a_1 e^{-\tau/\tau_2} + a_2 e^{-\tau/\tau_3}
\end{equation}

where $\lambda_i = 1/\tau_i, i\in \{1,2,3\}$ are the measured transition rates for the two NV system. We observe the 3 leading exponents that should be visible for two indistinguishable NV centers exhibiting superradiance [refer to Sec.~\ref{Case 2: Two NV Centers without interaction}]. Further, we compare the amplitudes of $g^{(2)}(0)$ for $P_1$ and $P_4$ via pulsed $g^{(2)}(\tau)$ measurements [see Fig. S3 of the Supplementary Information for details]. We see that the pulsed measurements are consistent with the $g^{(2)}(0)$ of the continuous-wave measurements. This indicates cooperative emission from the two emitters. The $g^{2}$ for cooperative emission from two indistinguishable NVs has a stark difference from that of two distinguishable NVs emitting incoherently (see Fig. S4 of Supplementary Information). 
From the fit, we extract the various transition rates. We comment on the extracted transition rates for the single and two indistinguishable emitter systems in Sec.~\ref{Transition Rates between various Energy Levels}.

\section{Theory of Intensity Correlations for Superradiance}

\subsection{Dynamical Master Equation}

        In the following, we will provide the master equation of $n$ NV centers interacting with an electric field. Next, we will derive formulas for the second-order correlation function for one NV center and two NV centers.
        %
        % ...
        %

\subsubsection{Dynamical Master Equation for $n$ NV centers}

The evolution of the density operator is given by the dynamical master equation~\cite{breuer2002theory}:

 \begin{align}
     \frac{d\hat{\rho}}{dt} 
     &= \frac{i}{\hbar} [\hat{\rho},\hat{H}_0] - \sum_{i,j}\sum_{\omega} \gamma_{ij}(\omega) (\hat{\sigma}_j(\omega)\hat{\rho}\hat{\sigma}^\dagger_i(\omega) \nonumber \\
     &- \{\hat{\sigma}^\dagger_i(\omega)\hat{\sigma}_j(\omega), \hat{\rho}\}) 
     \equiv \hat{\mathcal{L}} \hat \rho 
     \label{MasterEq1}
 \end{align}
 where $\hat{H}_0$ denotes the Hamiltonian of the reversible part of the dynamics,
 $\gamma_{ij}(\omega_{eg})$ and $\gamma_{ij}(\omega)$, $\omega \in \{\omega_{ig}, \omega_{ei}\}$  denote the 
 decay rates
 for, respectively, photonic and phononic transitions and $\hat{\sigma}_i(\omega)$ is a jump operator between pairs of states differing by an energy $\hbar \omega$.
 Finally, the indices $i$ and $j$ denote the label of an emitter in the system, i.e., $i, j \in \{1,2\}$.
 The value of $\gamma(\omega_{eg})$ depends on the state $\rho_B$ of the electric field interacting with the NV centers and the dipole moment matrix elements between energy levels. 

 The electric field is a thermal field, displaced in the narrow band around a wavelength $\lambda_\alpha = 520$ nm by a continuous-wave laser:
 \begin{align}
 \rho_B = \bigotimes_{\vec k,\lambda} 
\hat D(\alpha(\vec k,\lambda)) (1-e^{-\beta\hbar\omega_{k}})e^{-\beta \hbar \omega_{k} b^\dagger_{\vec k,\lambda} b_{\vec k,\lambda}}
\hat D(\alpha(\vec k, \lambda))^\dagger
\end{align}
where $\alpha(\vec k, \lambda)$ is supported in a narrow neighbourhood of $k_\alpha = 2\pi / 520$ nm. The value of displacement $\alpha(\vec k, \lambda)$ is related to the power spectral density of the laser: $\mathcal{P}(\omega) = dP/d\omega = \hbar \omega |\alpha(\vec k, \lambda)|^2$. 
 
The radiative decay rates $\gamma_{ij}(\omega)$ are:

\begin{align}
    \gamma_{ij}(\omega) 
    &= 
    \frac{2\omega^3 |d|^2}{3\pi\hbar\epsilon_0 c^3} f(x_{ij},\theta_{ij}) (1+N(\omega)) 
    \nonumber \\
    &+ 
    \frac{2\omega^3 |d|^2}{(4\pi)^2\hbar\epsilon_0 c^3}
     |\alpha(\omega)|^2
\end{align}

where $\omega, d$ correspond to the frequency of the radiative transition and dipole moment matrix element of the transition, respectively.  Also, $f(x_{ij},\theta_{ij}) = j_0 (x_{ij}) + P_2(\cos\theta_{ij})j_2 (x_{ij})$, $x_{ij} = \frac{\omega_k}{c}|r_i - r_j|$ and $\cos \theta_{ij} = \frac{|\vec d.(\vec r_i - \vec r_j)|^2}{d^2 |\vec r_i - \vec r_j|^2}$. $j_0$ and $P_2$ are the Bessel function of 0-th order and Legendre polynomial of 2nd order, respectively (see Supplementary Information). 

\subsection{Second-order autocorrelation function}
In terms of the jump operators $\{\hat \sigma, \hat \sigma^\dagger\}$, the second-order autocorrelation function is given by~\cite{bradac2017room, cygorek2023signatures}:

\begin{align}\label{g2_ladder_ops}
    g^{(2)}(\tau) = 
    \frac{
    \langle \hat{\sigma}^\dagger(0)\hat{\sigma}^\dagger(\tau)\hat{\sigma}(\tau)\hat{\sigma}(0)\rangle_{\rho_\infty}
    }
    {\langle\hat{\sigma}^\dagger(0)\hat{\sigma}(0)\rangle_{\rho_\infty}
    \langle\hat{\sigma}^\dagger(\tau)\hat{\sigma}(\tau)\rangle_{\rho_\infty}
    }
\end{align}
where $\tau$ is given as the time delay in detection between two detections after the beam-splitter in the experimental setup. 
The expected values are calculated with respect to stationary state $\rho_\infty$ of the system dynamics (the kernel of the respective Lindbladian):
\begin{align}\label{g2_generator}
    g^{(2)}(\tau) 
    &= 
    \frac{
    tr( \hat{\sigma}^\dagger(0)\hat{\sigma}^\dagger(\tau)\hat{\sigma}(\tau)\hat{\sigma}(0) \rho_\infty)
    }
    {
    tr ( \hat{\sigma}^\dagger(0)\hat{\sigma}(0) \rho_\infty)^2
    }
    \nonumber \\
    &=
    \frac{
    tr( e^{\mathcal{L}^\dagger \tau}(\hat{\sigma}^\dagger\hat{\sigma})\hat{\sigma} \rho_\infty \hat{\sigma}^\dagger)
    }
    {
    tr ( \hat{\sigma}^\dagger\hat{\sigma} \rho_\infty)^2
    }
\end{align}

In case of one NV center, the jump operator $\hat \sigma$ is $\ketbra{g}{e}$ and 
in case of two NV centers, the jump operator is $\ketbra{ge+eg}{gg} / \sqrt{2} +\ketbra{ee}{eg+ge})$. 
In both cases the operator $\hat{\sigma}^\dagger\hat{\sigma}$ is diagonal.

From the properties of Davies operators~\cite{breuer2002theory}, we know that the 
Linbladian $\mathcal{L}$ has a block structure $\mathcal{L}_{diag} \oplus \mathcal{L}_{offdiag}$, where $\mathcal{L}_{diag}$ act in the subspace of diagonal matrices, and $\mathcal{L}_{offdiag}$ in its orthogonal complement
(diagonal and off-diagonal part of the density matrix $\hat{\rho}$ evolve independently). Hence the dynamics of $\hat{\sigma}^\dagger\hat{\sigma}$ is always governed by $\mathcal{L}_{diag}$.

\subsection{One NV center}\label{One NV Center}

The system hamiltonian is now: $\hat H_0 = \sum_{r\in\{e,i,g\}} E_r\ketbra{r}$. The frequencies are $\omega_{lm} = (E_m - E_l) / \hbar$. 
The master equation preserves the trace, hence one eigenvalue is $0$ (the corresponding left eigenvector is $[1,1,1] \equiv \mathbbm{1}^T$). The stationary state 
$\rho_\infty$
is the kernel (right eigenvector to the eigenvalue $0$) of $\mathcal{L}_{diag}$ (see Supplementary Information for the exact form of $\rho_\infty$) and the propagator $\exp(\mathcal{L}t)$ is:
\begin{equation}\label{g2_single_emitter}
\exp(\mathcal{L}t) = \rho_\infty \mathbbm{1}^T + e^{\lambda_1 t} \ketbra{u_1}{v_1} +  e^{\lambda_2 t} \ketbra{u_2}{v_2}, 
\end{equation}
where $\{\rho_\infty, u_1, u_2\}$ and $\{\mathbbm{1}, v_1, v_2\}$ are left and right eigenbases of $\mathcal{L}_{diag}$ respectively, and $0,\lambda_1,\lambda_2$ are its eigenvalues, which (under a simplifying assumption, that $\gamma(\omega_{ig})+\gamma(\omega_{ei}) << \gamma(\omega_{eg}$), $\gamma(\omega_{}) = \gamma(\omega_{}) = 0$, reasonable for the experiment temperature around $300$ K) are equal: 
\begin{align}
    \lambda_1 &= \gamma(\omega_{eg}) + \gamma(\omega_{ge}) \label{l1_oneNV}\\
    \lambda_2 &= \gamma(\omega_{ei}) + \frac{\gamma(\omega_{ig})\gamma(\omega_{ge})}{\gamma(\omega_{eg}) + \gamma(\omega_{ge})} \label{l2_oneNV}
\end{align}

As the denominator of the $g^{(2)}$ function is time independent (equal to $\bra{e} \rho_\infty \ket{e}^2$, see (\ref{g2_generator})), the correlation function is a combination of a constant and two exponential decays.

\subsection{Two NV centers 
} \label{Case 2: Two NV Centers without interaction}

Let us consider first a system of two, non-interacting NV centers. The system Hamiltonian is now $\hat H_0 = \sum_{r} E_r (\ketbra{r}\otimes I + I\otimes \ketbra{r})$. The jump operators for the first and second NV center at transition frequency $\omega_{lm}$ are $\hat \sigma_1 (\omega_{lm}) = \ketbra{l}{m} \otimes I$ and $\hat \sigma_2 (\omega_{lm}) = I \otimes \ketbra{l}{m}$. 
Since the NV centers are indistinguishable and has integer spin, we are interested in the dynamics of the symmetric subspace of $\mathbb{C}^3 \otimes \mathbb{C}^3$, which is 6-dimensional. We obtain a 6-dimensional quantum Pauli equation for populations in corresponding Dicke states. Further, for superradiance, we consider $\gamma_{ij} = \Gamma_{ij} \forall i,j$~\cite{breuer2002theory}.

Again, the denominator of the $g^{(2)}(\tau)$ function is independent of $\tau$. The correlation function is a combination of 5 exponents and one constant. While we were working on this article, a recent study was reported by Qu et al.~\cite{qu2024superradiance}, where the numerics depicting the $g^{(2)}(\tau)$ function have been performed for superradiance in 2 NV centers. Our experimental results also match with the expected behaviour of the $g^{(2)}(\tau)$ function from the study by Qu et al. The situation does not change qualitatively if we add an interaction between the NV centers (we will still have 5 exponents).

Also, in~\cite{qu2024superradiance} the results for short correlation time scale (not covered in our experiment) show quick damped oscillations of $g^{(2)}(\tau)$ function. This indicates that a pair of eigenvalues of $\mathcal{L}_{diag}$ are conjugated complex numbers. The remaining 3 real eigenvalues will suffice to fit the $g^{(2)}$ function to the experimental data, proving the existence of (at least) two NV centers in the sample.  

\begin{figure*}[htp]
        \centering
\includegraphics[width=0.7\textwidth]{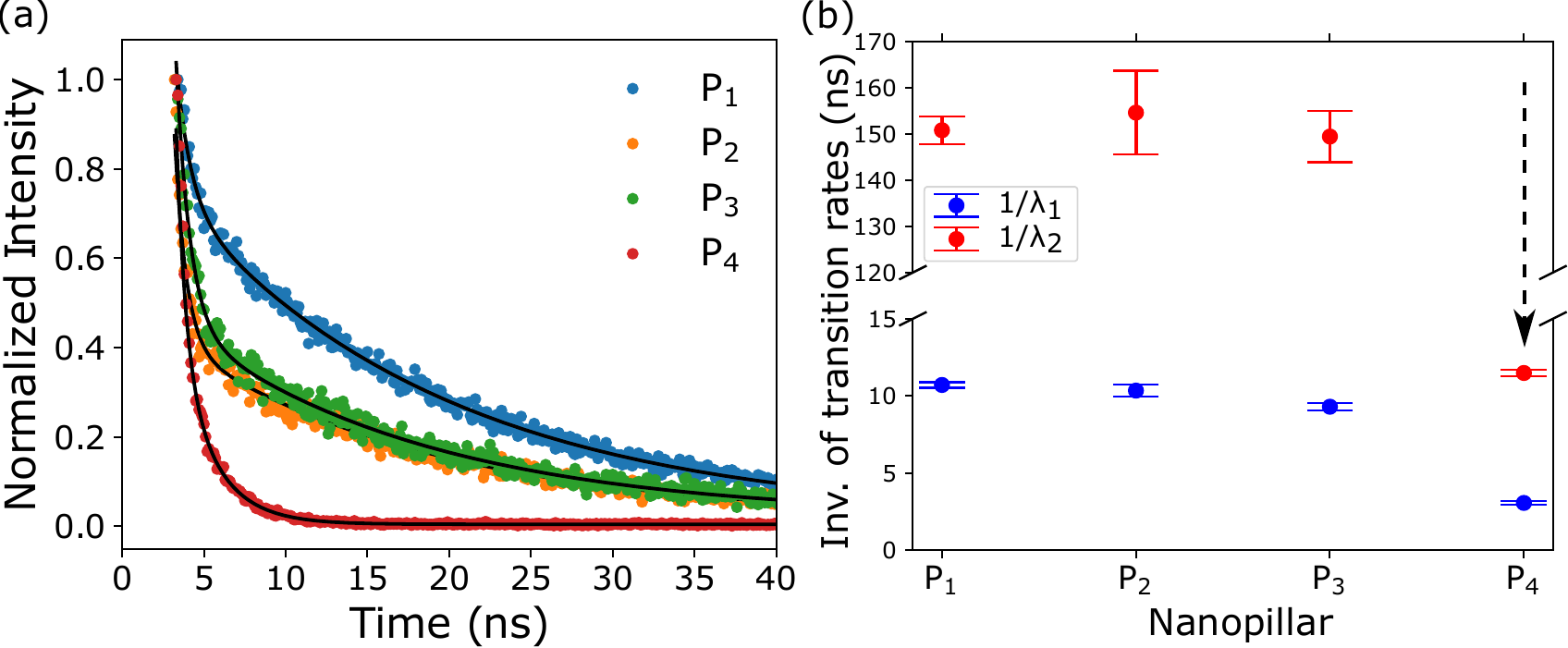}
        \caption{\textbf{Comparison of time constants of transitions from $\{P_1, \dots, P_4\}$: } (a) Normalized time-resolved PL intensity (measured: dots; solid line for fitting with Eqs.~\ref{lifetime_biexp}) for four nanopillars containing NV centers. It is plotted as a function of time delay $\tau$ in detecting coincidences. (b) Extracted time constants ($\tau_1, \tau_2$: inverses of transition rates) from $g^{(2)}(\tau)$. The time constants are associated with transitions $\ket{e}, \ket{i} \to \ket{g}$ and the total radiative and non-radiative transitions for single NVs ($P_1, \dots, P_3$), and two-NV system ($P_4$), respectively.
        $\tau_1$ and $\tau_2$ are shown in blue and red dots with error bars, respectively. The dashed arrow indicates the radiative and non-radiative decay time-scales for $P_4$ which is reduced by a factor of $\sim 6$ than the other emitters
        .
        }
\label{fig:LifetimeCompare}
    \end{figure*}

\begin{figure*}[ht]
        \centering
    \includegraphics[width=0.8\textwidth]{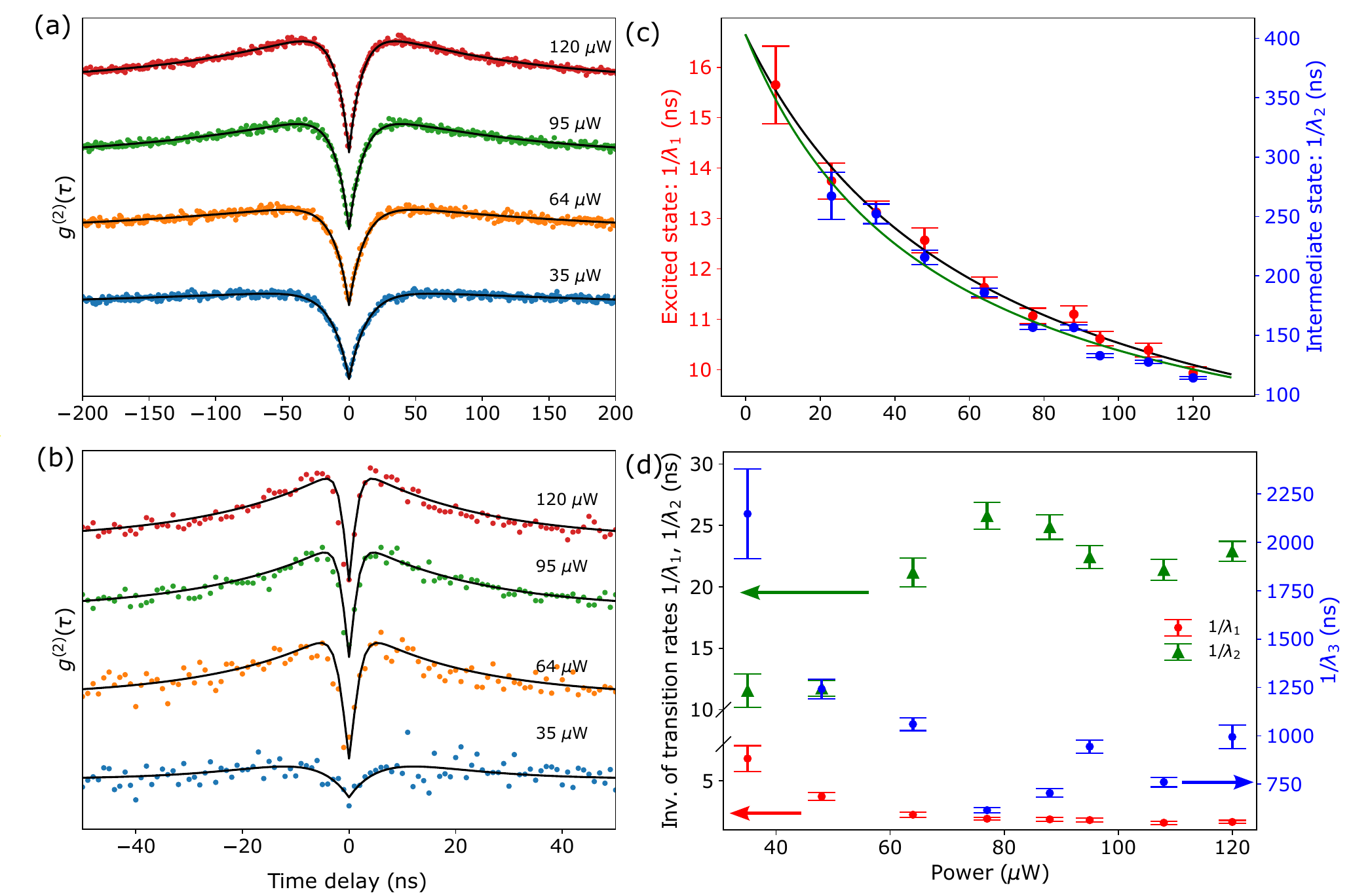}
        \caption{\textbf{Time constants of the transitions with pump power: } Measured $g^{(2)}(\tau)$ functions at various pump powers are shown for (a) $P_1$ and (b) $P_4$ [measured: dots; fitting: solid line]. The extracted parameters from the $g^{(2)}$ function for (c) $P_1:$ $1/\lambda_1 = \tau_1$: red dots, $1/\lambda_2 = \tau_2$: blue dots and (d) $P_4:$ $1/\lambda_1 = \tau_1$: blue dots, $1/\lambda_3 = \tau_1$: green triangles and $1/\lambda_3 = \tau_1$: red dots - right axis. For $P_1$, the solid lines show the fittings of $1/\lambda_1$ and $1/\lambda_2$ (with Eqs.~\ref{l1_oneNV} and~\ref{l2_oneNV}) as a function of pump power.
        }
        \label{fig:PowG2_Lifetimes}
    \end{figure*}

\section{Transition Rates between various Energy Levels}\label{Transition Rates between various Energy Levels}

\subsection{Comparison of radiative and non-radiative transition rates}

We perform a start-stop measurement of the lifetime (inverse of decay rate) for the radiative transition of the single NV centers. We use a pulsed excitation source at a 20 MHz repetition rate with a pulse width of $\approx 200$ ps. 
The time-resolved photoluminescence (PL) intensity decays (see Fig.~\ref{fig:LifetimeCompare} (a)) are fitted using the following function:

\begin{equation}\label{lifetime_biexp}
    I(t) = y_0+a_1 e^{-t/t_1}+a_2e^{-t/t_2}
\end{equation}

where $y_0$ is the offset, $a_1, a_2$ are the amplitudes of the exponents with time constants $t_1, t_2$, respectively.
The decay occurring on timescales $< 1$ ns can be attributed to the background within the emitting crystal~\cite{schietinger2009plasmon, dastidar2024structural}. Thus, the principle time constant arising due to PL decay from the emitters $P_i, (i \in \{1, 2, 3, 4\})$ are $16.6 \pm 0.24\text{ ns}, 15.7 \pm 0.36\text{ ns}, 14.7 \pm 0.38\text{ ns and } 2.05 \pm 0.03$ ns, respectively. 
For the single emitters $P_1, ..., P_3$, the PL lifetime is close to the reported value for a single NV center, i.e., $\approx 12$ ns, as expected. However, for $P_4$, we observe a drastic reduction by 6 times, further indicating cooperative emission from the two NV centers. We also show the comparison of the decay constants ($\tau_1, \tau_2$)
 extracted from the $g^{(2)}$ function in Fig.~\ref{fig:LifetimeCompare} (b). We see both $\tau_1$ and $\tau_2$ reduce to $\sim 1/6$th of the single emitters for the two NV centers, indicating an interaction between the emitters. We study the transitions further using power-dependent measurements. 

\subsection{Power-dependent Measurements}

Through the second-order correlation function, one can determine the individual decay rates involved in the system. This allows one to gain a better understanding of the transitions within the energy-level structure. For this purpose, we measure the second-order correlation function for $\{P_1, \dots, P_4\}$ as a function of excitation power with the HBT interferometer. For brevity and comparison of single and two emitter systems, the $g^{(2)}(\tau)$ function and the extracted transition rates for $P_1$ and $P_4$ as a function of power are shown in Figs.~\ref{fig:PowG2_Lifetimes} (a)-(d). The $g^{(2)}(\tau)$ functions as a function of power for the pillars $P_2, P_3$ are shown in the Supplementary information (Fig. S4). The measured and fitted $g^{(2)}(\tau)$ curves for some of the powers for the single emitter ($P_1$) and two-emitter ($P_4$) systems are shown in Fig.~\ref{fig:PowG2_Lifetimes} (a) - (b), respectively. It can be seen as the power increases, there is an increase in the amplitude of the bunched behaviour, arising due to the shelving of the emitter in the metastable state~\cite{treussart2001photon,aharonovich2010photophysics}. 

 The characteristic time scales (inverse of transition rates: $\tau_1 = 1/\lambda_1$ and $\tau_2 = 1/\lambda_2$) are plotted with excitation power for $P_1$ and $P_4$ in Fig.~\ref{fig:PowG2_Lifetimes} (c) and (d), respectively. Recall Eqs.~\ref{l1_oneNV} and~\ref{l2_oneNV}. For $P_1$ the expected hyperbolic behaviour w.r.t $r_{eg}$ (optical pumping) of $\tau_i (\equiv 1/\lambda_i, i\in\{1,2\})$ is observed as a function of power, which we recover from Eqs.~\ref{l1_oneNV} -~\ref{l2_oneNV} [see Fig.~\ref{fig:PowG2_Lifetimes} (c)]. In the limit of zero pump power, the spontaneous emission lifetimes are recovered (from fitting Eqs.~\ref{l1_oneNV} -~\ref{l2_oneNV}), as $\approx 16$ ns (excited state) and $\approx 400$ ns (metastable state) for $P_1$, as expected for single NV centers~\cite{storteboom2015lifetime}. 

    For $P_4$, $\tau_1$ and $\tau_2$ follow a hyperbolic trend with the excitation power. However, $\tau_3$ approaches a saturated value [see Fig.~\ref{fig:PowG2_Lifetimes} (d)]. The hyperbolic behaviour of $\tau_1$ is expected as the radiative transition rate ($1/\lambda_1$) is directly proportional to the optical pumping rate. This is also confirmed using the direct start-stop measurement. Further, the decay constants ($\tau_1, \tau_2, \tau_3$) represent the combinations of the time evolution of the populations of various energy levels. For 2 NV centers exhibiting superradiance, as discussed, we must restrict to the symmetric subspace: $\{\ket{gg}, \ket{ii}, \ket{ee}, (\ket{eg}+\ket{ge})/\sqrt{2}, (\ket{ei}+\ket{ie})/\sqrt{2}, (\ket{ig}+\ket{gi})/\sqrt{2}\}$ ($6\times 6$ operator space). Now, $\tau_i (\equiv 1/\lambda_i), i \in \{1,2,3\}$ are inverses of leading eigenvalues of the $5\times 5$ block of the Lindbladian, governing the dynamics of two indistinguishable NV centers exhibiting superradiance. The variation of the eigenvalues w.r.t. the optical pumping rate (an element of the matrix) is shown by the power-dependent studies. Thus, such a behaviour can be expected for eigenvalues of $5\times 5$ stochastic matrices. However, obtaining the exact analytical expression for the said is very challenging. 
    % ...  

\section{Quantum Random Number Generation}

To study one application of two-emitter systems, we discuss the possibility of quantum random number generation from these sources. We predict a better performance of the system of two NV centers due to the narrower width of the dip in the $g^{(2)}$ function. This is a consequence of the Wiener-Khinchin theorem, i.e., the power spectral density of a narrower $g^{(2)}$ function is closer to that of white noise (most random source).
The emitted PL from each of $\{P_1, \dots, P_4\}$ is split using a symmetric beam-splitter, thereby using the inherent randomness of the "which-path" choice faced by each photon in the stream of PL.

\subsubsection{Statistical Tests for Determining Randomness}
We record time-tagged photon arrivals across two single-photon detectors (SPD) to obtain a random binary sequence (see Methods section for details). We obtain a bit-generation rate of 540 kHz, 325 kHz, 760 kHz, and 170 kHz from the nanopillars $P_1, P_2, P_3$ and $P_4$, respectively, as we restrict to lower excitation powers (discussed in Sec.~\ref{Entropy}). The unbiased sequences are tested using a Python implementation of the NIST Statistical Test Suite, which is generally used to evaluate a binary sequence's randomness.

 

\begin{figure}[ht]
        \centering
    \includegraphics[width=0.5
\textwidth]{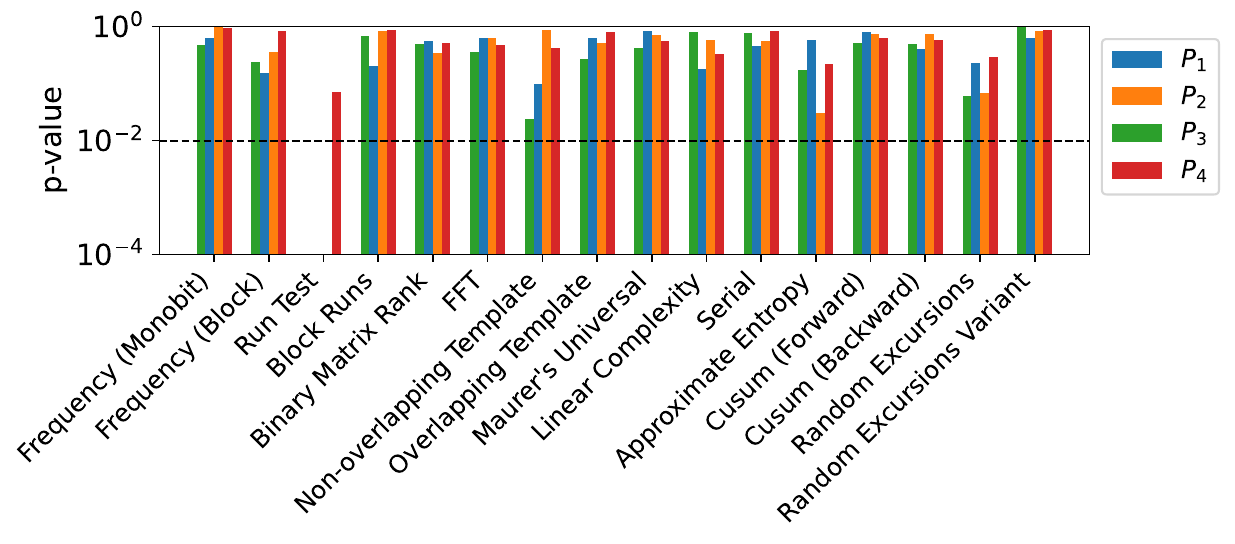}
        \caption{\textbf{Comparison of quantum random number generation: }NIST test results for the random bit sequences generated from $P_1, \dots, P_4$. The bits generated from
        $P_4$ pass all the tests with p-values $> 0.01$ with a higher proportion, indicating stronger reliability of randomness from such two coupled emitter systems.}
        \label{fig:QRNG}
\end{figure}

As shown in Fig.~\ref{fig:QRNG}, the random number sequences generated for nanopillars $P_1, P_2, P_3$ fail the runs test whereas that for $P_4$ passes all the subtests, determined by the $p-$value, i.e., $p > 0.01$. The runs test determines whether uninterrupted sequences of identical bits occur as expected for an ideal random sequence of the same length as the measured sequence. Since, $P_1,\dots,P_3$ have a wider $g^{(2)}$ function, their power spectral density is narrow. This leads to the generation of alternating bit sequences which fail the runs test. Whereas, $P_4$, has short-range correlations due to narrower a $g^{(2)}$ dip, in contrast, and thus passes the runs test. We have observed that on repeated trials of data acquisition from the emitters, $P_4$ consistently show randomness by passing all the NIST subtests.

\subsubsection{Entropy of the source \& non-idealities of detection}\label{Entropy}


Here, we consider the effect of non-idealities of the measurement setup on random number generation. Asymmetricity ($R \neq T$) of the beam-splitter ($R, T$), detection efficiencies ($\eta_A, \eta_B$) and the dead times ($\tau^A, \tau^B$) of the detectors (denoted by $A$ and $B$) add noise to the randomness of the generated binary bit sequence. 
Also, dead times act as correction factors relating the count rate to the actual photon flux onto the detector. For low incoming photon count rates and dark counts, this factor is close to unity~\cite{chen2019single}. Thus, we restrict our measurements to low excitation power and comparatively low count rates. 


To quantify the randomness of the sources, we calculate the extractable entropy, i.e., conditional min-entropy ($H_\infty$) of the binary sequence using~\cite{chen2019single}:

\begin{equation}\label{MinEntropy}
    H_{\infty} (X|Y) = -\log_2\Big(\sum_y p(y)\max\{p(x|y)\}\Big)
\end{equation}

where $X, Y$ are two events, $x, y$ are the subsequent random bits and $p(y), p(x|y)$ are the probabilities for occurrence of $y$ and conditional occurrence of $y$ if $x$ occurs.
For our case, $\{X, Y\} \in\{0,1\}$, thus Eq.~\ref{MinEntropy} can be written as:
\begin{align}\label{MinEntropy2}
    H_{\infty}(X|Y) = &-\log_2\Big(p(0)\max\{p(0|0),p(1|0)\} \nonumber \\ &+ p(1)\max\{p(0|1),p(1|1)\}\Big)
\end{align}

Using the parameters for our detectors: $\eta_A = \eta_B = 40 \%, \tau^A = 77.9$ ns, $\tau^B = 74.7$ ns and beam-splitter: $R = 0.55, T = 0.45$ we obtain, $p(A) = \{0.6097, 0.6098, 0.6095, 0.7098\}$ for emitters $\{P_1, P_2, P_3, P_4\}$, respectively (see detailed derivation in Supplementary Information). Further, $p(A)$ is the maximum value for function $\{p(A), 1-p(A), 2p(AB), 1-2p(AB)\}$. Thus, for the pillars $\{P_1, P_2, P_3, P_4\}$, we extract $H_\infty(X|Y) = \{0.75126, 0.75129, 0.751201, 0.77081\}$, respectively with $H_\infty(X|Y) > 0 \forall$ emitters and $H_\infty(X|Y)$ being maximum for $P_4$. Since conditional min-entropy acts as a metric for randomness, we can conclude that all pillars generate random sequences with $P_4$ having a higher randomness per quantum bit.

\section{Conclusion}

We experimentally show optical superradiance from two NV centers in a diamond nanopillar. We observe a $g^{(2)}(0) > 0.5 \to 1$ and drastic reductions in the lifetimes of the excited and intermediate states of the system, which are considered as the signature of cooperative effects from a system of emitters. Further, we analytically solve the Lindblad master equation for two indistinguishable NV centers interacting with a common electric field, in the regime of superradiance. We qualitatively show that the second-order autocorrelation function ($g^{(2)}(\tau)$) for such a system must contain three exponential terms, due to the additional symmetries in the system. We establish the functional form of  Finally, we show an application of our system of emitters in the context of random number generation. The two-emitter system showing superradiant behaviour can produce reliable random number sequences with a generation rate of $\sim$200 kHz at low pump powers. 

%

\section{Methods}

\subsection{Sample}\label{Sample}
The sample (7 keV irradiated, single NV arrays from QZabre Ltd.) consists of a diamond membrane with nanopillar arrays, containing on an average low concentration of NVs. The collection efficiency is 10$\times$ compared to bulk diamond. The nanopillar diameter measured was $\sim 500$ nm.

\subsection{Confocal microscope \& HBT setup with Continuous-Wave and Pulsed Excitation}

To characterize the sample, we set up a home-built confocal microscope integrated with the Hanbury-Brown Twiss (HBT) interferometer, to perform optical measurements on the sample. We use a diode laser (PicoQuant, LDH-IB-520-B) emitting light at 520 nm wavelength, with two operational modes, continuous-wave (CW) \& pulsed for photoluminescence (PL) lifetime and intensity correlation measurements, respectively. The time-resolved start-stop measurements of PL intensity were done at 20 MHz repetition rate, with $\sim 200$ ps pulse width and average power is $\sim 70 \mu$W. Also, the pulsed coincidence measurements were done in the same settings as the time-resolved intensity measurements. The cw coincidence ($g^{(2)}$) measurements were done at 300 $\mu$W power. To obtain a diffraction-limited spot at the sample, we ensure that the fundamental emission mode is outputted from the laser by using a single-mode optical fiber. The diverging light from the fiber is expanded and collimated using a biconvex lens. This beam is then directed towards the aperture of a 100X, infinity-corrected objective lens with a numerical aperture (NA) of 0.9, after passing it through a dichroic mirror. The spot size of the beam is calculated using~\cite{meinhart2003theory}
$ d_\infty = 1.22\lambda_{pump}\sqrt{\Big(\frac{n}{\text{NA}}\Big)^2-1}$, where $\lambda_{pump} = 520$ nm (excitation wavelength), $n = 1$ (refractive index of the medium of beam propagation) and NA$ = 0.9$. Using the above equation, the beam spot size is $320$ nm. 

The sample is placed on a stack of three piezo-driven nanopositioners (attoCube, ECSx3030 and ECSz3030) to move in $(x,y,z)$ directions. From the sample, emitted PL is collected by the objective and transmitted through the dichroic mirror and a long pass filter to pass wavelengths $>$650 nm. To block the out-of-focus PL (wavelength: $\lambda_{PL}$), the diameter of the pinhole is estimated using $d_{pinhole} = d_{\infty}M\frac{\lambda_{PL}}{\lambda_{pump}}$, where $M$ is the magnification of the objective. Using the above formula and the measured beam spot size $d_{\infty} \approx 570$ nm, we obtain the required pinhole diameter, $d_{pinhole} \approx 90$ $\mu$m.

For detection, we use free-space, single photon avalanche diode (SPAD) detectors (PDM series, PD-100-CTB) with 100 $\mu$m sensor diameter. Thus, the active area of the SPADs acts as a pinhole to block out-of-focus light. To perform intensity correlation measurements (HBT setup), the PL is focused onto two SPADs by passing it through a non-polarizing, symmetric beam splitter. The signal is collected from the SPADs using a Time-Correlated Single Photon Counter (TCSPC: PicoQuant Multiharp 150P). The coincidences are measured with a time delay implemented electronically and are restricted by the TCSPC's resolution, 5 ps. Further, since the emitters are embedded in sub-micron size regions of the sample, we need to focus the excitation light on a particular region using white-light imaging. For this, we focus a white-light beam onto the sample using the objective and obtain the image of the sample onto a camera sensor. This is performed before the actual experiment to locate the region of interest for excitation.

\subsection{Data Acquisition Techniques}

We performed PL lifetime imaging and coincidence imaging to locate the nanopillar, containing emitters satisfying the conditions for superradiance. During spatial PL intensity mapping, for selective pixels (above a threshold intensity set at the beginning of the acquisition), both time-resolved intensity and coincidences are collected. The real-time acquisition software was home-built using Python. The binwidth was set to be 1 ns for better signal-to-noise ratio of the measurements. We scanned multiple pillars on the sample and identified the ones containing pillars with emitters.

We discuss a random binary sequence generator. 
The PL from each emitter-system of interest is inputted to one input mode of a beam-splitter and the other input arm is blocked (describing the vacuum state $\ket{0}$). Each photon in the stream of photons of the emission will randomly get transmitted or reflected based on the splitting ratio of the beam-splitter. Thus, the inherent quantumness of the "which-path" concept is used as the principle for random number generation. Clicks in one SPD are recorded as 0 and the other is recorded as 1, which depends on the symmetricity of the beam-splitter. The sequence of raw random bits is post-processed using von Neumann's de-biasing procedure, to extract unbiased random numbers for our study~\cite{luo2020quantum}. Using this, we eliminate some co-dependence of two adjacent bits. For every pair of generated bits, 00 and 11 are discarded, and 01 and 10 are replaced by 0 and 1, respectively. We perform this debiasing protocol in real-time using our data acquisition program.

\end{document}


\title{Supplementary Information: Signatures of superradiance in intensity correlation measurements in a two-emitter solid-state system}
\author{Madhura Ghosh Dastidar}
 
    \affiliation{Quantum Center of Excellence for Diamond and Emerging Materials (QuCenDiEM) Group, Department of Physics, Indian Institute of Technology Madras, Chennai 600036, India}

	\author{Aprameyan Desikan}

	\affiliation{Department of Physical Sciences, Indian Institute of Science Education \& Research Mohali Sector 81 SAS Nagar, Manauli PO 140306 Punjab, India}

	\author{Gniewomir Sarbicki}
 
	\affiliation{Institute of Physics, Faculty of Physics, Astronomy and Informatics, Nicolaus Copernicus University, Grudzi\c adzka 5/7, 87-100 Toru\'n, Poland}

	\author{Vidya Praveen Bhallamudi$^{\ast}$}

    \affiliation{Quantum Center of Excellence for Diamond and Emerging Materials (QuCenDiEM) Group, Department of Physics, Indian Institute of Technology Madras, Chennai 600036, India}
\maketitle



\tableofcontents

\section{Energy levels of NV center}\label{Energy levels of NV center}

The single-hole Hamiltonian takes in the $\{\sigma_1, \sigma_2, \sigma_3, \sigma_N\}$ basis a most general form allowed by the symmetry of the defect~\cite{maze2011properties}:

\begin{align}
    V &= V_N\ketbra{\sigma_N} + \sum_i \{V_C \ketbra{\sigma_i} + (h_N \ketbra{\sigma_i}{\sigma_N} +h.c.)\} + \sum_{i>j} h_C\ketbra{\sigma_i}{\sigma_j} + h.c. 
    &= \begin{pmatrix}
        V_C & h_C & h_C & h_N \\
        h_C & V_C & h_C & h_N \\
        h_C & h_C & V_C & h_N \\
        h^*_N & h^*_N & h^*_N & V_N \\
    \end{pmatrix} 
    \label{electron_ion_matrix}
\end{align}

As shown in Appendix \ref{first_hybr}, the eigenvalues of the Hamiltonian are (in the increasing order for a hole):
\begin{align}
E_{xy} = & V_C - h_C, \qquad \text{2-fold degenerated} \\
E_1 = &\frac {V_C+V_N}2 + h_C - \sqrt{
\left( \frac {V_C+V_N}2 + h_C \right)^2 + 3h_N^2
} \\
E_2 = & \frac {V_C+V_N}2 + h_C + \sqrt{
\left( \frac {V_C+V_N}2 + h_C \right)^2 + 3h_N^2
},
\end{align}
and the corresponding eigenvectors are:
\begin{align}\label{single_electron_orbitals}
    \ket{x} = \frac{1}{\sqrt{6}}(2\sigma_1 - \sigma_2 - \sigma_3) \\
    \ket{y} = \frac{1}{\sqrt{2}}(\sigma_2 - \sigma_3)\label{single_electron_orbitals3} \\
    \ket{1} = \frac{\alpha_+}{\sqrt{3}}(\sigma_1+\sigma_2+\sigma_3) + \alpha_- \sigma_N \\
    \ket{2} = -\frac{\alpha_-}{\sqrt{3}}(\sigma_1+\sigma_2+\sigma_3) + \alpha_+ \sigma_N,
\end{align}
where $\alpha_\pm$ are given Eqs. \ref{evecs}. As the energy $E_2$ is highest, since now we will restrict ourselves to Hilbert space spanned by vectors $\ket{x}, \ket{y}$ and $\ket{1}$. 

From the single electron orbitals, we find the 2-hole states and estimate their energy levels. 

The orthogonal basis generator formula \cite{weissbluth2012atoms} given by:
\begin{align}\label{2Hole_Wavefn1}
     \Psi_r 
    &= \frac{l_r}{h} \sum_e \chi^r_e (R_e \otimes R_e) (\phi_i \otimes \phi_j) 
\end{align}
returns the component of $\phi_i \otimes \phi_j$ in the $r$-th irreducible representation (irrep) of the $C_{3v}$ group of symmetry of the defect. Here $h=6$ denotes the rank of the group, $l_r$ denotes the dimension of the $r$-th irrep, $R_e$ denotes the group operation and $\chi^r_e$ denotes its character in the $r$-th representation (see Appendix~\ref{Energy levels of NV center} for details). The $C_{3v}$ group consists of the identity element, rotations by $\pm \frac{2\pi}{3}$, and reflections in 3 vertical planes passing through each of the carbon atoms.
The group has 3 irreducible representations denoted as $A_1$, $A_2$ and $E_1$ (trivial, sign, and faithful, respectively), of dimensions $1, 1$ and $2$ respectively.
The characters of subsequent symmetry operations in these representations are:
\begin{align}
    \chi^{A_1} = & \{1,1,1,1,1,1\} \label{char_A1}\\
    \chi^{A_2} = & \{1,1,1,-1,-1,-1\} \label{char_A2} \\
    \chi^{E_1} = & \{2,-1,-1,0,0,0\} \label{char_E1}
\end{align}

Applying Eq.~\ref{2Hole_Wavefn1}, one obtains the following basis of the 9 two-hole orbitals (see Appendix \ref{second_hybr} for details), grouped by the irreps they belong to. Further, the irreps are divided into their symmetric and antisymmetric sector, and the corresponding energies are the following:

\begin{table}[htp]
\begin{tabular}{|c|c|c|}
\hline
irrep & state & energy 
\\ \hline \hline
$A_1^+$ &
$\frac 1{\sqrt{2}}(\ket{xx} + \ket{yy})$ 
&
$2E_{xy}$
\\
& $\ket{11}$ 
&
$2E_1$ 
\\
\hline
$A_2^-$ & $\frac 1{\sqrt{2}}(\ket{xy} - \ket{yx})$
& $2E_{xy}$
\\
\hline
$E_1^+$ 
&
$\frac{1}{\sqrt{2}}(\ket{xy} - \ket{yx})$
&
$2E_{xy}$ \\
&
$\frac{1}{\sqrt{2}}(\ket{xx} + \ket{yy})$
&
$2E_{xy}$
\\ 
&
$\frac 1{\sqrt{2}}(\ket{1j}+\ket{j1})$, $j \in \{x,y\}$
&
$E_1 + E_{xy}$ \\
\hline
$E_1^-$ & 
$\frac 1{\sqrt{2}}(\ket{1j}-\ket{j1})$, $j \in \{x,y\}$
& $E_1 + E_{xy}$ 
\\ \hline
\end{tabular}
\label{tab:orbitals}
\caption{9-dimensional space of 2-hole orbitals divided into symmetric and antisymmetric sectors of subsequent representations of $C_{3v}$ group.}
\end{table}

These states are states of two holes with a two-particle Hamiltonian without the Coulomb interaction term. Considering this interaction as a perturbation, it does not mix states of different representations. Further, it neither mixes the symmetric and antisymmetric sectors. In the above 2-hole orbital basis, the matrix of perturbation with elements defined as $\bra{\Psi_i(\vec r_1,\vec r_2)}V(|\vec r_1 - \vec r_2|)\ket{\Psi_j(\vec r_1,\vec r_2)}$ is block-diagonal with block sizes 2, 1, 4 and 2, respectively. This perturbation removes degeneracy and slightly raises the energies of various levels. It affects the symmetric orbitals more than the antisymmetric orbitals.

The spin space of two holes is spanned by three symmetric spin functions $\{ \frac 1{\sqrt 2} (\uparrow \otimes \uparrow \pm \downarrow \otimes \downarrow), \frac 1{\sqrt{2}} ( \uparrow \otimes \downarrow + \downarrow \otimes \uparrow )\}$ and one antisymmetric spin function $\frac 1{\sqrt{2}} ( \uparrow \otimes \downarrow - \downarrow \otimes \uparrow )$. They belong to the following representations:
\begin{center}
\begin{tabular}{|c|c|c|}
\hline
irrep & state
\\ \hline
\hline
$A_1^-$ & 
$\frac 1{\sqrt 2}(\ket{\uparrow\downarrow}-\ket{\downarrow\uparrow})$
\\ \hline
$A_2^+$ & $\frac 1{\sqrt 2}(\ket{\uparrow\downarrow}+\ket{\downarrow\uparrow})$ \\ \hline
$E_1^+$ & 
\begin{tabular}{@{}c@{}}
$\frac 1{\sqrt 2}(\ket{\uparrow\uparrow}+\ket{\downarrow\downarrow})$
\\
$\frac 1{\sqrt 2}(\ket{\uparrow\uparrow}-\ket{\downarrow\downarrow})$
\end{tabular} 
\\ \hline
\end{tabular}
\end{center}

Now we combine the antisymmetric orbitals with symmetric spin functions and vice versa, using a projection formula including the spin part as well. This allows us to build 15 antisymmetric spin-orbitals of two holes and identify the representations to which they belong (see Appendix \ref{spin-orbitals} for details). We obtain the following result: 


\noindent
\begin{center}
    \begin{tabular}{|c|c|c|}
\hline
irrep & state & SP
\\ \hline
\hline
$A_1$
&
$
\left. \begin{array}{r}
\frac 1{\sqrt 2}(\ket{xx} + \ket{yy})
\\
\ket{11}
\end{array} \right\} \otimes \frac 1{\sqrt 2}
(\ket{\uparrow\downarrow} - \ket{\downarrow\uparrow})
$
&
$A_1^+$
\\
&
$\frac 12 (\ket{xy} - \ket{yx}) 
\otimes (
\ket{\uparrow\downarrow} + \ket{\downarrow\uparrow}
)
$ 
&
$A_2^-$
\\
&
$
\frac 12
(
(\ket{1-}-\ket{-1})\otimes\ket{\uparrow\uparrow}
-
(\ket{1+}-\ket{+1})\otimes\ket{\downarrow\downarrow}
)
$
&
$E_1^-$
\\ \hline
$A_2$
&
$
\frac 12
(
(\ket{1-}-\ket{-1})\otimes\ket{\uparrow\uparrow}
+
(\ket{1+}-\ket{+1})\otimes\ket{\downarrow\downarrow}
)
$
&
$E_1^-$
\\ \hline
&
$
\frac 12 (\ket{xy} - \ket{yx})
\otimes
\ket{\uparrow\uparrow} \pm \ket{\downarrow\downarrow}
$
&  
$A_2^-$
\\
$E_1$
&
$
\left. \begin{array}{l} 
\ket{1j}+\ket{j1}, \quad j \in \{x,y\} \\
\ket{xy}+\ket{yx} \\
\ket{xx}-\ket{yy}
\end{array} \right\}
\otimes \frac 12 (\ket{\uparrow\downarrow} - \ket{\downarrow\uparrow})
$
&
$E_1^+$
\\
&
$
\frac 12 (\ket{1j}-\ket{j1}) \otimes (\ket{\uparrow\downarrow} + \ket{\downarrow\uparrow})
, \quad j \in \{x,y\}
$
&
$E_1^-$
\\
&
$
\frac 12
(
(\ket{1+}-\ket{+1})\otimes\ket{\uparrow\uparrow}
\pm
(\ket{1-}-\ket{-1})\otimes\ket{\downarrow\downarrow}
)
$
&
$E_1^-$
\\ \hline
\end{tabular}
\end{center}

\noindent where 
$\ket{\pm} = (\ket{x} \pm i \ket{y})/\sqrt{2}$. 
The last column specifies the representation type of the spatial part. It is important since the radiative transitions are allowed between states having a non-vanishing matrix element of a component of the dipole moment. The $x$ and $y$ components of the dipole moment transform as $E_1^+$ representation, while the $z$ component transforms as $A_2^+$ representation. Hence the pairs of states giving the non-vanishing matrix element are of types: $(E_1, E_1)$, $(A_2, A_1)$ for the $z$ component, while for perpendicular components one of the states has to be of $E_1$ type. Both states have to have the same permutational symmetry.

From now on, we will refer to these states as $A_1(i)$, $A_2(i)$, and $E_1(i)$, where the numbering is in the order of appearance in the respecting row in the table above.

At this stage, we have three degenerated spin triplets emerging from three antisymmetric spatial functions. This degeneracy will be removed by the next perturbation $-$ the spin-orbit interaction. The $k$-th hole energy operator of this interaction is given by:
\begin{align}
H_{SO}(k) 
& = \frac{(g_s - 1) \mu_B}{2 e m_e c^2} \left( \vec \nabla_k \cross \vec p \right) \cdot \vec \sigma
\nonumber \\
& \approx
\frac{\mu_B}{2 e m_e c^2} \left( \vec \nabla_k \cross \vec p \right) \cdot \vec \sigma
\nonumber \\
& \stackrel{df}{=} 
\frac{\mu_B}{2 e m_e c^2} \vec \Omega \cdot \vec \sigma
,
\end{align}
where $\mu_B = \hbar e / (2m_e c)$ is the Bohr magneton, $g_s$ is the electron-spin g-factor and $\vec \sigma$ is the vector of Pauli matrices. 
The $x,y$ components of the operator $\vec \Omega$ transforms as $E$ representation, while the $z$ component transforms as the $A_2$ representation. Taking it into account, we find the expression for the components of $\vec \Omega$ in the $\{\ket{x},\ket{y},\ket{1}\}$ basis:
\begin{align}
    \Omega_x = \left[ \begin{array}{ccc}
        0 & 0 & a \\ 0 & 0 & 0 \\ a^* & 0 & 0 
    \end{array} \right],
    \Omega_y = \left[ \begin{array}{ccc}
        0 & 0 & 0 \\ 0 & 0 & a \\ 0 & a^* & 0 
    \end{array} \right],
    \Omega_z = \left[ \begin{array}{ccc}
        0 & b & 0 \\ b & 0 & 0 \\ 0 & 0 & 0 
    \end{array} \right]
\end{align}

The two-hole spin-orbit operator is then of the form:
\begin{align}
H_{SO} = \frac{\mu_B}{2 e m_e c^2} \sum_{i \in \{x,y,z\}} \Big( & \Omega_i \otimes I_3 \otimes \sigma_i \otimes I_2  + 
I_3 \otimes \Omega_i \otimes I_2 \otimes \sigma_i
\Big)
\end{align}

In the basis $\{ A_1(i), A_2(i), E_1(i)\}$ It's matrix is block-diagonal. In the subspace of states transforming as $A_1$ the block has the form:
\begin{equation}
    H_{SO}(A_1) = \left[\begin{matrix}0 & 0 & - 2 i \lambda_z & - \sqrt{2} \lambda_\perp\\0 & 0 & 0 & 2 \lambda_\perp\\2 i \lambda_z & 0 & 0 & - \sqrt{2} i \lambda_\perp\\- \sqrt{2} \lambda_\perp & 2 \lambda_\perp & \sqrt{2} i \lambda_\perp & \lambda_z\end{matrix}\right],
\end{equation}
where $\lambda_z = \frac{\mu_B}{2 e m_e c^2} b$ and $\lambda_\perp = \frac{\mu_B}{2 e m_e c^2} a$ 
The next block $H_SO(A_2) = \lambda_z$ is one-dimensional.
The subspace of states $\{E_1(i)\}$ decomposes into a direct sum of four subspaces invariant w.r. to spin-orbit interaction. The corresponding blocks of $H_{SO}$ are:
\begin{align}
    H_{SO}(E_1(1), E_1(3), E_1(8)) = 
    \left[\begin{matrix}0 & - i \lambda_\perp & \lambda_\perp\\i \lambda_\perp & 0 & - i \lambda_z\\\lambda_\perp & i \lambda_z & 0\end{matrix}\right]
    \\
    H_{SO}(E_1(2), E_1(4), E_1(7)) = 
    \left[\begin{matrix}0 & - \lambda_\perp & i \lambda_\perp\\- \lambda_\perp & 0 & i \lambda_z\\- i \lambda_\perp & - i \lambda_z & 0\end{matrix}\right]
    \\
    H_{SO}(E_1(5), E_1(9)) =
    \left[\begin{matrix}0 & - \sqrt{2} i \lambda_\perp\\\sqrt{2} i \lambda_\perp & - \lambda_z\end{matrix}\right]
    \\
    H_{SO}(E_1(6), E_1(10)) =
    \left[\begin{matrix}0 & - \sqrt{2} \lambda_\perp\\- \sqrt{2} \lambda_\perp & - \lambda_z\end{matrix}\right]
\end{align}
Four states: $A_1(4), A_2(1), E_1(9), E_1(10)$ from spin triplets $E_1^-$ gain energy correction $\pm \lambda_z$ and their degeneracy is partially removed. The non-zero off-diagonal terms show possible non-radiative transitions between states of different permutational symmetry of the spatial part of wave function. 

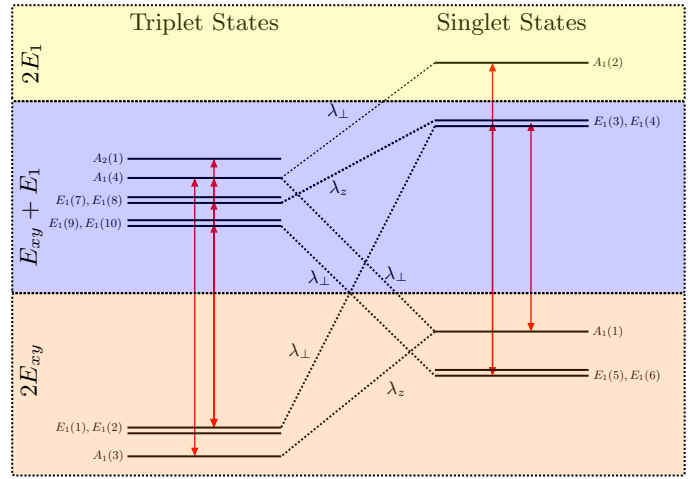
\begin{figure}[htp]
\begin{adjustbox}{width=0.5\linewidth}
 \begin{tikzpicture}
  \draw[dashed, line width=2pt] (17, 12) -- (9, 6);
  \draw[dashed, line width=3pt] (9, 6) -- (17, -2.);
  \draw[dashed, line width=3pt] (17, -2.) -- (9, -8.5);
  \draw[dashed, line width=3pt] (9, 4.7) -- (17, 9);
  \draw[dashed, line width=3pt] (9, 4.7) -- (17, 9);
  \draw[dashed, line width=3pt] (17, 8.7) -- (9, -7.);
  \draw[dashed, line width=3pt] (9, 3.5) -- (17, -4.3);

   \node[scale=4] at (5, 14) {Triplet States};
   \node[scale=4] at (21, 14) {Singlet States};
  \node[scale=3] at (15, -5) {$\lambda_z$};
  \node[scale=3] at (15, 1) {$\lambda_\perp$};
  \node[scale=3] at (12, 9.5) {$\lambda_\perp$};
  \node[scale=3] at (10, -3) {$\lambda_\perp$};
  \node[scale=3] at (12, 5.5) {$\lambda_z$};
  \node[scale=3] at (11, 0.8) {$\lambda_\perp$};

  \draw[>=triangle 45, <->, red, line width=2pt]  (4.5, -8.5) -- (4.5, 6);
  \draw[>=triangle 45, <->, red, line width=2pt]  (5.5, -7) -- (5.5, 4.8);
  \draw[>=triangle 45, <->, red, line width=2pt]  (5.5, -7) -- (5.5, 3.6);
  \draw[>=triangle 45, <->, red, line width=2pt]  (5.5, -7) -- (5.5, 6);
  \draw[>=triangle 45, <->, red, line width=2pt]  (5.5, -7) -- (5.5, 7);
  
   \draw[>=triangle 45, <->, red, line width=2pt]  (20, -4.3) -- (20, 8.9);
   \draw[>=triangle 45, <->, red, line width=2pt]  (20, -4.3) -- (20, 12);
   \draw[>=triangle 45, <->, red, line width=2pt]  (22, -2) -- (22, 8.9);

  \draw[thick, line width=3pt] (17, 12) -- (25, 12);
\node[scale=2] at (26, 12) {$A_1(2)$};
  \draw[thick, line width=3pt] (17, 9) -- (25, 9);
  \draw[thick, line width=3pt] (17, 8.7) -- (25, 8.7);
  
  \node[scale=2] at (27, 8.9) {$E_1(3), E_1(4)$};

  \draw[thick, line width=3pt] (1, 7) -- (9, 7);
  \node[scale=2] at (0., 7) {$A_2(1)$};
  \draw[thick, line width=3pt] (1, 6.) -- (9, 6.);
  \node[scale=2] at (0., 6) {$A_1(4)$};
  \draw[thick, line width=3pt] (1, 5.) -- (9, 5.);
  \draw[thick, line width=3pt] (1, 4.7) -- (9, 4.7);
  \draw[thick, line width=3pt] (1, 3.8) -- (9, 3.8);
  \node[scale=2] at (-1, 4.8) {$E_1(7), E_1(8)$};
  \draw[thick, line width=3pt] (1, 3.5) -- (9, 3.5);
  \node[scale=2] at (-1.2, 3.6) {$E_1(9), E_1(10)$};

  \draw[thick, line width=3pt] (17, -2.) -- (25, -2.);
\node[scale=2] at (26, -2) {$A_1(1)$};
  
  \draw[thick, line width=3pt] (17, -4.) -- (25, -4.);
  
  \draw[thick, line width=3pt] (17, -4.3) -- (25, -4.3);
  \node[scale=2] at (27, -4.3) {$E_1(5), E_1(6)$};

  \draw[thick, line width=3pt] (1, -7.) -- (9, -7.);
  \node[scale=2] at (-1, -7) {$E_1(1), E_1(2)$};
  \draw[thick, line width=3pt] (1, -7.3) -- (9, -7.3);
  \node[scale=2] at (0, -8.5) {$A_1(3)$};
  \draw[thick, line width=3pt] (1, -8.5) -- (9, -8.5);

  \draw[rounded corners, dashed, line width = 3pt, fill=orange, fill opacity=0.2] (-5, -9.5) rectangle (30, 0) {};
  \node[scale=4, rotate=90] at (-4, -4) {$2E_{xy}$};
  \draw[rounded corners, dashed, line width = 3pt, fill=blue, fill opacity=0.2] (-5, 0) rectangle (30, 10) {};
  \node[scale=4, rotate=90] at (-4, 4) {$E_{xy}+E_1$};
  \draw[rounded corners, dashed, line width = 3pt, fill=yellow, fill opacity=0.2] (-5, 10) rectangle (30, 15) {};
  \node[scale=4, rotate=90] at (-4, 12) {$2E_1$};
\end{tikzpicture}
\end{adjustbox}
\caption{Energy diagram of lowest states of an $NV^-$ center.}
\label{fig:states}
\end{figure}

The third perturbation considered in the literature, weaker then the previous ones, is the spin-spin interaction. It removes the energy degeneracy in the pair $A_1(4),A_2(1)$ and separates the doublet $E_1(1),E_1(2)$ from $A_2(1)$ \cite{maze2011properties}. We summarize the energy levels of the NV center in Fig.~\ref{fig:states} showing the states and their energies, possible radiative transitions (red arrows) and non-radiative transitions (dotted lines) mediated by the spin-orbit coupling.

We  treat the NV center as the following three-level system:
\begin{align}
        \ket{g} = A_1(3) &= \frac{1}{2}(\ket{xy} - \ket{yx}) \otimes (\ket{\uparrow\downarrow} + \ket{\downarrow\uparrow}) \\
        \ket{i} = A_1(1) &= \frac{1}{2}(\ket{xx} + \ket{yy}) \otimes (\ket{\uparrow\downarrow} - \ket{\downarrow\uparrow}) \\
        \ket{e} = A_1(4) &= \frac 12(
            (\ket{1-}-\ket{-1})\otimes\ket{\uparrow\uparrow}
            \nonumber \\
            & -
            (\ket{1+}-\ket{+1})\otimes\ket{\downarrow\downarrow}
        )
\end{align}
This is because the closed cycle of excitation and emission for our experiment occurs in the states with $A_1$, namely, ($A_1(3), A_1(1), A_1(4)$) representation.

\section{Residuals of the $g^{(2)}$ fitting}

\begin{figure}[htp]
    \centering
    \includegraphics[width=0.8\linewidth]{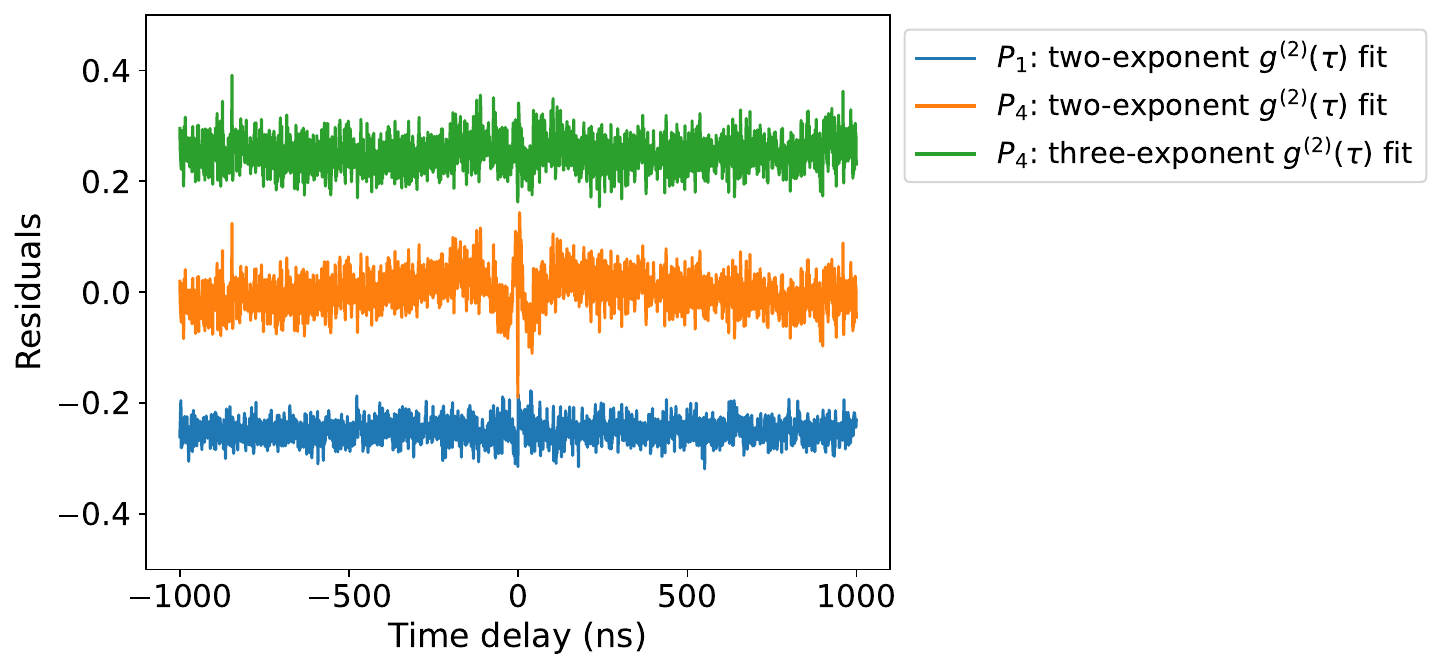}
    \caption{Comparison of residuals for $g^{(2)}$ measured for $P_1$ and $P_4$ with the different fitting functions (two exponents: Eq. (5) and three-exponents: Eq. (6) of the main manuscript). A manual offset is added for clear visibility.}
    \label{fig:residuals}
\end{figure}

In Fig.~\ref{fig:residuals}, we show the residuals of the fits for $g^{(2)}$ function for $P_1$ and $P_4$. We use the $g^{(2)}(\tau)$ functions given by Eqs. (5) and (6) in the main manuscript are described as functions with two and three exponents, respectively. We observe a clear non-random behaviour when $g^{(2)}$ is fitted with the two-exponent $g^{(2)}$ function, characteristic for a single NV center. This behaviour gets randomized when we use $g^{(2)}$ function with three-exponents, making it a better fit for the measured data from $P_4$, i.e., nanopillar containing two emitters.

\section{Radiative decay rate} \label{app:gamma}

 To derive the radiative 
 rates of thermalization ($\gamma(\omega_{eg})$) for our experimental setup, we consider the radiation field as the bath and a system of single or two emitters. The interaction Hamiltonian in the dipole approximation is given as:

 \begin{align}
     H_I = - \sum^n_{j = 1} \hat {\vec D}_j \cdot \vec E (\vec r_j)
 \end{align}
where 
\begin{equation}
\hat {\vec D}_j =
e^+ (\vec r_j^{(1)} \otimes I + I \otimes \vec r_j^{(2)}) 
= 
\langle e_j | ( r_j^{(1)} \otimes I + I \otimes \vec r_j^{(2)} ) | g_j \rangle | e_j \rangle \langle g_j | + \text{h.c.} 
\equiv 
\vec d | e_j \rangle \langle g_j | + \text{h.c.}
\end{equation} 
is the dipole operator of the $j$-th NV center
(observe, that the $j$-the center dipole operator has only one non-vanishing matrix element corresponding to the radiative transition $|e\rangle \to |g\rangle$ (obviously the second one, conjugated corresponds to transition $|g\rangle \to |e\rangle$), and the vector $\vec d$ of matrix elements of components of dipole operator is the same for all NV centers)
and
    \begin{align}\label{electric_field}
        \vec{E}(\vec{r_i}) =  
        \sum_{\lambda, \vec k} \sqrt{\frac{\hbar\omega_k}{2V\epsilon_0}}
        \vec{e}_{\lambda,\vec k}(
        b_{\lambda,\vec k}e^{i\vec k \cdot \vec r_i}-b^\dagger_{\lambda, \vec k} e^{-i\vec k \cdot \vec r_i})
    \end{align}
is the electric field in the $j$-th NV center.

To calculate the thermalization rates we calculate the one-sided Fourier transform of the bath correlation function:
\begin{align}
    \Gamma_{ij}
    (\omega) = \frac 1{\hbar^2} \int^\infty_0 dt \exp{i\omega t} \langle \vec{d}^* \cdot\vec{E}^\dagger (\vec{r}_i, t)\vec{d} \cdot \vec{E}(\vec{r}_j, 0) \rangle_{\rho_B}.
    \label{bath_correlation_func}
\end{align}
Since now $\omega$ means the frequency of the radiative transition.
Using the Eq.~\ref{electric_field}, we have:
\begin{align}
    \Gamma_{ij}(\omega) 
    &
    = 
    \frac 1{\hbar^2}
    \sum_{k,k'}\sum_{\lambda,\lambda'} 
    \sqrt{\frac{\hbar\omega_k}{2V\epsilon_0}}\sqrt{\frac{\hbar\omega_k'}{2V\epsilon_0}} 
    \vec d^* \cdot \vec{e}_{\lambda, \vec k}
    \vec d \cdot \vec{e}_{\lambda',\vec k'}
    \int^\infty_0 dt e^{i\omega t} 
    \Big[
    -\langle b_{\lambda, \vec k}b_{\lambda',\vec k'} \rangle_{\rho_B}
    e^{i\vec k .(\vec r_i+\vec r_j)}e^{-i\omega_k t} 
    \nonumber \\
    &
    + 
    \langle b^\dagger_{\lambda, \vec k}b_{\lambda',\vec k'} \rangle_{\rho_B} 
    e^{-i\vec k .(\vec r_i-\vec r_j)}e^{i\omega_k t}  
    + 
    \langle b_{\lambda, \vec k}b^\dagger_{\lambda',\vec k'} \rangle_{\rho_B}
    e^{i\vec k .(\vec r_i-\vec r_j)}e^{-i\omega_k t} 
    -
    \langle b^\dagger_{\lambda, \vec k}b^\dagger_{\lambda',\vec k'} \rangle_{\rho_B}
    e^{-i\vec k .(\vec r_i+\vec r_j)e^{i\omega_k t}}
    \Big]
    \label{S_1}
\end{align}
Using the fact that:
\begin{align}\label{mean_vals}
    \langle b_\lambda(\vec{k})b_{\lambda'}(\vec{k'}) \rangle_{\rho_B}
    &= \langle b^\dagger_\lambda(\vec{k})b^\dagger_{\lambda'}(\vec{k'}) \rangle_{\rho_B}
    = 0 \\
    \langle b^\dagger_\lambda(\vec{k})b_{\lambda'}(\vec{k'}) \rangle_{\rho_B} 
    &= \delta_{kk'}\delta_{\lambda \lambda'} \bar{n}_{\vec k, \lambda}\\ 
    \langle b_\lambda(\vec{k})b^\dagger_{\lambda'}(\vec{k'}) \rangle_{\rho_B} 
    &= \delta_{kk'}\delta_{\lambda \lambda'}(1+\bar{n}_{\vec k, \lambda}),
\end{align}
where $\bar{n}_{\vec k, \lambda}$ is the mean number of photons in the mode. Now
the above simplifies to:
\begin{align}
    \Gamma_{ij}(\omega) = 
    \sum_{k,\lambda} 
    \frac{\omega_k}{2V\epsilon_0\hbar}
    |\vec d \cdot \vec{e}_{\lambda,\vec k}|^2
    \int^\infty_0 dt e^{i\omega t} 
    &\Big[ 
    \bar{n}_{\vec k, \lambda} 
    e^{-i\vec k .(\vec r_i-\vec r_j)}e^{i\omega_k t}
    +
    (1+\bar{n}_{\vec k, \lambda}) 
    e^{i\vec k .(\vec r_i-\vec r_j)}e^{-i\omega_k t} 
    \Big]
    \label{S_1}
\end{align}
The electromagnetic field is a room-temperature thermal field affected in mode $\lambda_\alpha = 532 nm$ by a laser, causing displacement of this mode: 
$\rho_B = \bigotimes_{\vec k,\lambda} 
\hat D(\alpha(\vec k,\lambda)) (1-e^{-\beta\hbar\omega_{k}})e^{-\beta \hbar \omega_{k} b^\dagger_{\vec k,\lambda} b_{\vec k,\lambda}}
\hat D(\alpha(\vec k, \lambda))^\dagger$,
where $\alpha(\vec k, \lambda)$ is supported in a narrow neighbourhood of $k_\alpha = 2\pi / 532$nm. Hence:
\begin{align}\label{n_avg}
    \bar{n}_{\vec k, \lambda}
    &=
    \langle b^\dagger_{\lambda, \vec k}b_{\lambda,\vec k}
    \rangle_{\rho_B}
    =
    tr(\hat{n}_{\lambda, \vec k}\rho_B) 
    = 
         N(\omega_k)+|\alpha(\vec k, \lambda)|^2
\end{align}
where $N(\omega) \equiv \frac{1}{e^{\beta\hbar\omega}-1}$ and the last equality follows from:
 $tr(\hat{a}^\dagger \hat a \hat{D}^\dagger(\alpha) \rho_{th} 
 \hat{D}(\alpha)) = tr(\hat{D}(\alpha)\hat{a}^\dagger \hat a \hat{D}^\dagger(\alpha)\rho_{th}) = tr((\hat{a}^\dagger - \alpha^*)(\hat{a} - \alpha)\rho_{th}) = tr(\hat{a}^\dagger\hat{a} \rho_{th}) - \alpha^*tr(\hat{a}\rho_{th}) -\alpha tr(\hat{a}^\dagger\rho_{th}) + |\alpha|^2 tr(\rho_{th}) = N(\omega_{\alpha}) + |\alpha|^2$ , where $\rho_{th} = (1-e^{-\beta\hbar\omega_k})e^{-\beta \hbar \omega_k b^\dagger_{\vec k,\lambda} b_{\vec k,\lambda}}$.


The term $\sum_\lambda |\vec d \cdot \vec{e}_{\lambda,\vec k}|^2$ is a square of the component of $\vec d$, which is perpendicular to $\vec k$:
$\sum_\lambda \vec d^\dagger \vec e_{\lambda,k} \vec e_{\lambda,k}^\dagger \vec d = \vec d^\dagger \Pi_{\vec k^\perp} \vec d = \vec d^\dagger (I - \vec k^T \vec k / k^2)\vec d$.

In the continuum limit of radiation modes, we have:
\begin{align}
    \frac{1}{V} \sum_{k} \rightarrow \int\frac{d^3k}{(2\pi)^3} = \frac{1}{(2\pi)^3}\int k^2 dk \int d\Omega = \frac{1}{(2\pi)^3}\int (\omega_k/ c)^2 d(\omega_k/c) \int d\Omega  = \frac{1}{(2\pi c)^3}\int^\infty_0 d\omega_k \omega^2_k \int d\Omega. 
\end{align}

The above observations let us to write:
\begin{align}
    \Gamma_{ij}(\omega) 
    =&
    \frac 1{2\hbar \epsilon_0 (2\pi c)^3}\int^\infty_0 d\omega_k \omega^3_k \int d\Omega
    (\vec d^*\cdot\vec d - |\vec d^* \cdot \vec k|^2 / k^2)
    \nonumber \\
    &
    \int^\infty_0 dt e^{i\omega t} 
    \Big
    [N(\omega_k)
    e^{-i\vec k \cdot (\vec r_i-\vec r_j)}
    e^{i(\omega_k+\omega) t} 
    +
    (1+N(\omega_k)) 
    e^{i\vec k  \cdot (\vec r_i-\vec r_j)}
    e^{-i(\omega_k-\omega) t} \Big]
    \nonumber \\
    &
    \frac 1{2\hbar\epsilon_0(2\pi c)^3}\int^\infty_0 d\omega_k \omega^3_k \int d\Omega
    (\vec d^*\cdot\vec d - |\vec d^* \cdot \vec k|^2 / k^2)
    \nonumber \\
    &
    \int^\infty_0 dt e^{i\omega t}
    |\alpha(\vec k, \lambda)|^2
    \Big[
    e^{-i\vec k \cdot (\vec r_i-\vec r_j)}
    e^{i(\omega_k+\omega) t} 
    + 
    e^{i\vec k  \cdot (\vec r_i-\vec r_j)}
    e^{-i(\omega_k-\omega) t} \Big]
    \label{S_2}
\end{align}

All NV centers lie in one plane and their axes are perpendicular to it. The laser also shines from the perpendicular direction, hence in the second summand we take $\vec k \cdot (\vec r_i - \vec r_j) = 0$. 
Moreover: as spatial parts of $\ket{g}$ and $\ket{e}$ transform as $A_2$ and $E_1$ respectively, $z$ as $A_2$ and $(x,y)$ as $E_1$, the $z$ component of vector $\vec d$ zeros, due to $A_2 \otimes A_2 \otimes E_1 \not \ni A_1$ and $\vec d \cdot \vec k = 0$. The factor $|d|^2$ goes outside the angle integral and the integral over $\Omega$ removes dependence on angle from the power spectrum - since now $\alpha$ is a function of $k$ or equivalently $\omega$. 

In the first summand the integral over angular degrees of freedom gives:
\begin{equation}
    \int d\Omega
    (\vec d^*\cdot\vec d - |\vec d^* \cdot \vec k|^2 / k^2) e^{\pm i\vec k  \cdot (\vec r_i-\vec r_j)}
    = \frac{8\pi}3 |\vec d|^2 [ j_0(x_{ij}) + P_2 \cos(\theta_{ij}) j_2(x_{ij}) ] \equiv 
    \frac{8\pi}3 |\vec d|^2
    f(x_{ij}, \theta_{ij}),
\end{equation}
where $j_0 (x) = \frac{\sin x}{x}, j_2(x) = (\frac{3}{x^3}-\frac{1}{x}) \sin x$ and $P_2(\cos \theta) = \frac{1}{2}(3\cos^2 \theta - 1)$, for  $x_{ij} = \frac{\omega_k}{c}|r_i - r_j|$ and $\cos \theta_{ij} = \frac{|\vec d \cdot (\vec r_i - \vec r_j)|^2}{d^2 |\vec r_i - \vec r_j|^2}$.

The time integrals will be derivated using: $\int^\infty_0 ds e^{-ixs} = \pi\delta(x) - i P\frac{1}{x}$ and $\int^\infty_0 ds e^{ixs} = \pi\delta(x) + i P\frac{1}{x}$, where $P$ is the Cauchy principal value. Now equation~(\ref{S_2}) takes the form:

\begin{align}
    \Gamma_{ij}(\omega) 
    &= 
    \frac{|d|^2}{6\pi^2 \hbar \epsilon_0 c^3} \Big[ 
    \int^\infty_0 d\omega_k \omega^3_k f(x_{ij},\theta_{ij}) \int^\infty_0 dt \Big((1+N(\omega_k)) e^{-i(\omega_k - \omega)t} + N(\omega_k) e^{i(\omega_k + \omega)t}\Big)
    \Big] \nonumber \\
     &+
     \frac {|d|^2}{2\hbar\epsilon_0(2\pi c)^3} \Big[
     \int^\infty_0 d\omega_k \omega^3_k 
     |\alpha(k,\lambda)|^2
     \int^\infty_0 dt \Big( e^{-i(\omega_k - \omega)t} + e^{i(\omega_k + \omega)t}\Big)
     \Big] \nonumber \\
     &= \frac{|d|^2}{6\pi^2 \hbar \epsilon_0 c^3}  
    \int^\infty_0 d\omega_k \omega^3_k f(x_{ij},\theta_{ij}) \Big[\pi(1+N(\omega_k))\delta(\omega_k - \omega) + \pi N(\omega_k) \delta(\omega_k + \omega)+iP \Big(\frac{1+N(\omega_k)}{\omega_k - \omega} - \frac{N(\omega_k)}{\omega_k + \omega}\Big)
    \Big] \nonumber \\
     &+
     \frac {|d|^2}{2\hbar\epsilon_0(2\pi c)^3} \Big[
     \int^\infty_0 d\omega_k \omega^3_k 
     |\alpha(k,\lambda)|^2
     \Big(\pi(\delta(\omega_k - \omega) + \delta(\omega_k + \omega)) - iP(\frac{1}{\omega_k + \omega} - \frac{1}{\omega_k-\omega})\Big)
     \Big]
    \label{S_4}
\end{align}

Using $N(-\omega) = -(1+N(\omega))$ and $\alpha(-\omega) = 0$ because the laser has a single mode and there is only a narrow frequency spread around $\omega$, we have:

\begin{align}
    \Gamma_{ij}(\omega) 
    &= 
    \frac{\omega^3 |d|^2}{3\pi\hbar\epsilon_0 c^3} f(x_{ij},\theta_{ij}) (1+N(\omega))
    +
    i\frac{|d|^2}{6\pi^2 \hbar \epsilon_0 c^3} f(x_{ij},\theta_{ij})\int^\infty_0 d\omega_k \omega^3_k  P \Big(\frac{1+N(\omega_k)}{\omega_k - \omega} - \frac{N(\omega_k)}{\omega_k + \omega}\Big)
     \nonumber \\
     &+
     \frac{\omega^3 |d|^2}{(4\pi)^2\hbar\epsilon_0 c^3}
     |\alpha(\omega)|^2
    +
    i \frac {|d|^2}{16\pi^3 \hbar c^3} \int^\infty_0 d\omega_k \omega^3_k |\alpha(\vec k,\lambda)|^2P\Big(\frac{1}{\omega_k - \omega} - \frac{1}{\omega_k + \omega}\Big) \\
      & \equiv \frac 12 \gamma_{ij}(\omega) + i S_{ij}(\omega) \nonumber
     \label{Gamma_rad_tot} 
\end{align}

Since the imaginary part ($S_{ij}(\omega)$) of Eq.~\ref{Gamma_rad_tot} will contribute to the Lamb shift and not affect the dynamics of the system, the radiative rates are governed by $\gamma_{ij}(\omega) = \Gamma_{ij}(\omega) + \Gamma_{ij}^*(\omega)$:

\begin{align}
    \gamma_{ij}(\omega) &= \frac{2 \omega^3 |d|^2}{3\pi\hbar\epsilon_0 c^3} f(x_{ij},\theta_{ij}) (1+N(\omega)) 
    +
    \frac{2\omega^3 |d|^2}{(4\pi)^2\hbar\epsilon_0 c^3}
     |\alpha(\omega)|^2
\end{align}

\section{One NV Center: Lindbladian}\label{One NV Center: Lindbladian}
The master equation in this case is:



\begin{widetext}

    \begin{align}
     \sum_{p,q} \dot{\rho}_{pq}\ketbra{p}{q} &= \frac{1}{i\hbar} \sum_r \sum_{p, q}E_r \rho_{pq} [\ketbra{r}, \ketbra{p}{q}] + \sum_{l\neq m} \gamma(\omega_{lm}) \sum_{p, q}\rho_{pq} \Big(\ketbra{l}{m} \ketbra{p}{q}\ketbra{m}{l} - \frac{1}{2}\{\ketbra{m},\ketbra{p}{q}\}\Big) \nonumber \\
     &= \frac{1}{i\hbar}\sum_{p, q} (E_p-E_q)\rho_{pq}\ketbra{p}{q} 
     + \delta_{pq} \sum_{l \ne p} \gamma(\omega_{lp})\ketbra{l}
     - \frac 12 \sum_{l \ne m} \gamma(\omega_{lm})
     \left(
     \sum_q \rho_{mq} \ketbra{m}{q}
     +
     \sum_p \rho_{pm} \ketbra{p}{m}
     \right)
 \label{Eq_NV1}
 \end{align}
\end{widetext}


    The equations for the diagonal part of density matrix (populations of the energy levels) is:

\begin{align}
    &\dot{\rho}_{pp} = 
    \sum_{l\neq p } \gamma(\omega_{lp})\rho_{ll}
    - \sum_{l \ne p} \gamma(\omega_{lp}) \rho_{pp},
\end{align} 
in the matrix notation:
\begin{align}
    \begin{pmatrix}
        \dot{\rho}_{gg} \\
        \dot{\rho}_{ii} \\
        \dot{\rho}_{ee}
    \end{pmatrix} &= \mathcal{L}_{diag} \begin{pmatrix}
        \rho_{gg} \\
        \rho_{ii} \\
        \rho_{ee}
    \end{pmatrix},
\end{align}
where
\begin{widetext}
    \begin{align}
    \mathcal{L}_{diag}
    &= \begin{pmatrix}
        -\gamma(\omega_{ig}) - \gamma(\omega_{eg}) &  \gamma(\omega_{ig}) & \gamma(\omega_{eg}) \\
        \gamma(\omega_{gi}) & -\gamma(\omega_{ei}) - \gamma(\omega_{gi}) & \gamma(\omega_{ei})\\   \gamma(\omega_{ge})  & \gamma(\omega_{ie}) & -\gamma(\omega_{ge}) - \gamma(\omega_{ie})
    \end{pmatrix}
\end{align}
\end{widetext}

The kernel of this Lindbladian is:

\begin{align}
\rho_\infty 
=& \left[ \begin{array}{c}
\gamma(\omega_{ei})\gamma(\omega_{ge}) + \gamma(\omega_{gi})\gamma(\omega_{ge}) + \gamma(\omega_{gi})\gamma(\omega_{ie})
\\
\gamma(\omega_{ge})\gamma(\omega_{ig}) + \gamma(\omega_{ie})\gamma(\omega_{ig}) + \gamma(\omega_{ie})\gamma(\omega_{eg})
\\
\gamma(\omega_{ig})\gamma(\omega_{ei}) + \gamma(\omega_{eg})\gamma(\omega_{ei}) + \gamma(\omega_{eg})\gamma(\omega_{gi})
\end{array} \right]
\nonumber \\
/\Big(& \gamma(\omega_{ei})\gamma(\omega_{ge}) + \gamma(\omega_{gi})\gamma(\omega_{ge}) + \gamma(\omega_{gi})\gamma(\omega_{ie})
\nonumber \\
&\gamma(\omega_{ge})\gamma(\omega_{ig}) + \gamma(\omega_{ie})\gamma(\omega_{ig}) + \gamma(\omega_{ie})\gamma(\omega_{eg})
\nonumber \\
&\gamma(\omega_{ig})\gamma(\omega_{ei}) + \gamma(\omega_{eg})\gamma(\omega_{ei}) + \gamma(\omega_{eg})\gamma(\omega_{gi}) \Big)
\end{align}

which is also the stationary state of the system.

\section{Pulsed $g^{(2)}$ Measurements}

\begin{figure}[htp]
    \centering
    \includegraphics[width=0.8\linewidth]{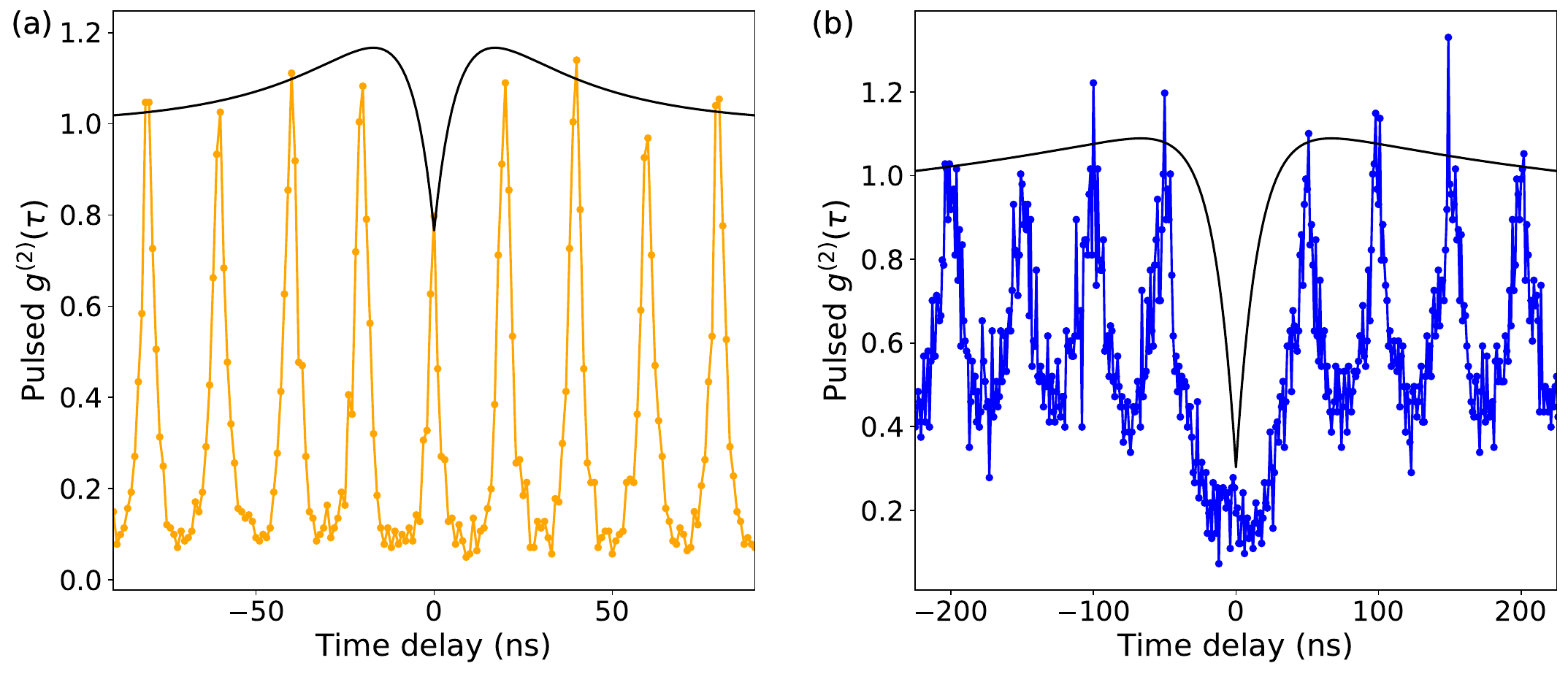}
    \caption{Measured (in pulsed laser mode, in dots) and envelope-fits (solid black line) for the $g^{(2)}$ functions for (a) $P_4$ and (b) $P_1$. Measurements are done at 50 MHz and 20 MHz repetition rates for $P_4$ and $P_1$ to resolve peaks distinctly.}
    \label{fig:Pulsed_g2}
\end{figure}

We perform pulsed measurements on the emitted PL from $P_4$ and $P_1$ to obtain the $g^{(2)}(\tau)$ functions second-order
correlation function [see Figs.~\ref{fig:Pulsed_g2} (a) and (b)]. Under pulsed excitation, the average power was kept at 35 $\mu$W for both emitter-systems. The peak at $\tau = 0 $ (zero time delay in detection) is the
probability of having more than one photon in the same
pulse. Thus, $P_1$ showing single-photon emission, has negligible peak at $\tau = 0$. Whereas, $P_4$ (two-emitter system) shows a distinct peak of $g^{(2)}(0) = 0.8$. Further, the envelopes of the peaks show the same behaviour as the intensity correlation measured under continuous-wave excitation (represented by solid black lines for $P_1$ and $P_4$).

\section{Intensity correlations for two distinguishable NV centers}

\begin{figure}[htp]
    \centering
    \includegraphics[width=0.5\linewidth]{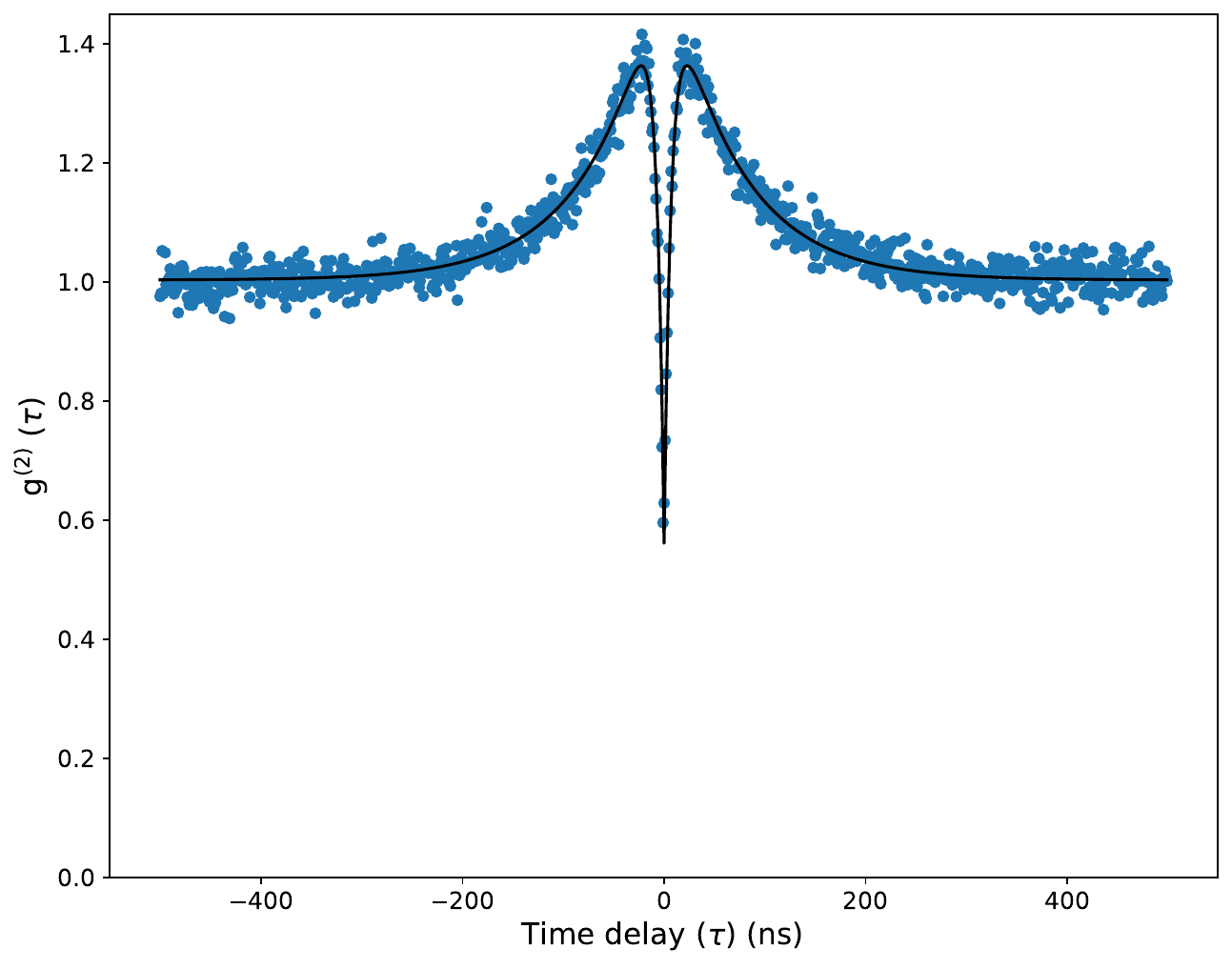}
    \caption{Measured (in dots) and fitted (solid line) $g^{(2)}(\tau)$ function for two-NV centers within the same nanopillar, which are distinguishable. Fitting is performed with Eq. (5) and $g^{(2)}(0) = 0.58$.}
    \label{fig:2emitter_g2}
\end{figure}

We show the $g^{(2)}(\tau)$ function for two NV centers, which are distinguishable, i.e., their dipole orientation is different [see Fig.~\ref{fig:2emitter_g2}]. We see that the functional form of the $g^{(2)}$ remains same as that of the single NV center. Further, $g^{(2)}(0) \simeq 0.58 \approx 0.5$, which is characteristic of a two-photon source~\cite{gerry2023introductory, worboys2020quantum, cygorek2023signatures}. There is no appearance of a third exponent, the characteristic two exponents defining the transitions from excited and intermediate states are enough to describe the $g^{(2)}$ for this case. Thus, this further highlights the difference in $g^{(2)}$ functions for distinguishable and indistinguishable NVs interacting with a common electric field within a volume less than the wavelength of excitation.

\section{Transition Rates for $P_2$ and $P_3$}

\begin{figure}[htp]
    \centering
    \includegraphics[width=0.8\linewidth]{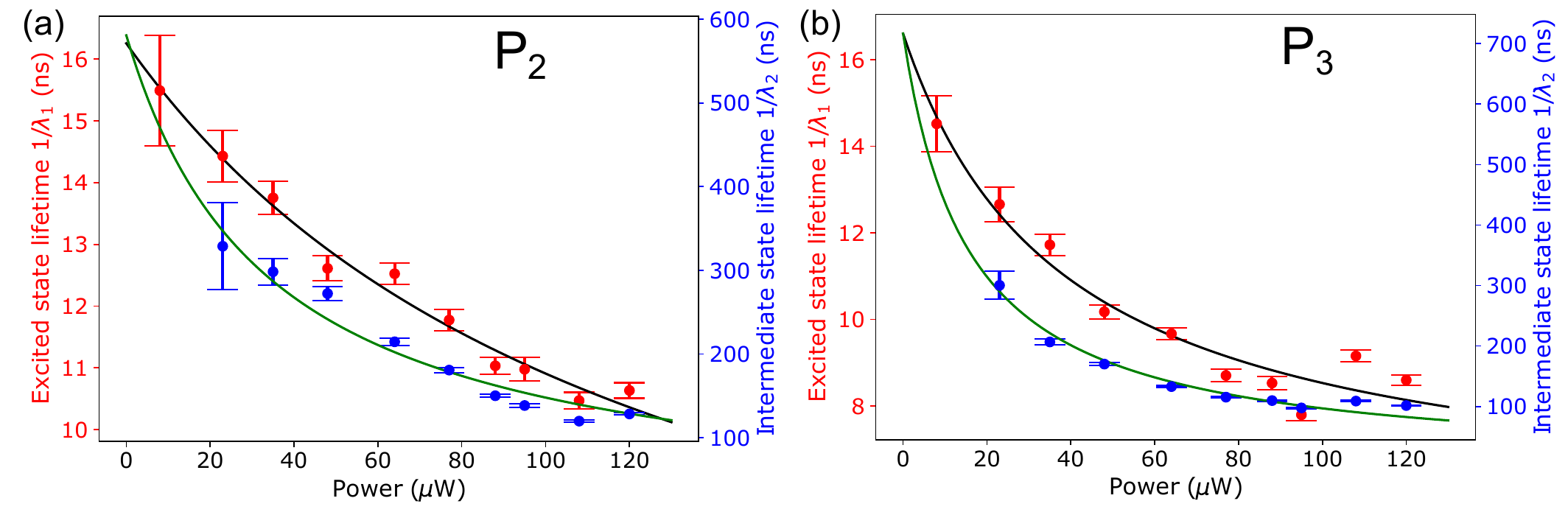}
    \caption{Extracted inverse of transition rates from excited and intermediate states to ground state for (a) $P_2$ and (b) $P_3$. Fitting is done to Eqs. (13) and (14) of the main text.}
    \label{fig:P2P3_compare_powg2}
\end{figure}

The power dependent $g^{(2)}$ measurements were conducted for nanopillars $P_2$ and $P_3$. From these measurements, we also extract the excited and intermediate state lifetimes $\tau_1 \equiv 1/\lambda_1, \tau_2 \equiv 1/\lambda_2$, respectively, for the other single emitters considered in our studies. This is done by performing fits to Eqs. (13) and (14). We observe for consistency that the transition rates of the single emitters ($P_2$ and $P_3$) fit well to the respective equations [see Figs.~\ref{fig:P2P3_compare_powg2} (a) and (b)]. Further, we extract the excited and intermediate state lifetimes as $\tau_1 = 16.22 \pm 0.26$ ns, $\tau_2 = 580\pm 14$ ns (for $P_2$) and $\tau_1 = 16.6 \pm 0.64$ ns, $\tau_2 = 716\pm 5$ ns (for $P_3$), close to the reported values in~\cite{storteboom2015lifetime}.



\section{Min-conditional Entropy for Randomness}\label{Min-conditional Entropy for Randomness}
Let us consider that the detector in the transmitted arm is A and that in the reflected arm is B. Therefore, the probability of the detector A (bit \textbf{1}) or detector B (bit \textbf{0}) clicking is given by $p(A) (\equiv p(0))$ or $p(B) (\equiv p(1))$, respectively. Similarly, the probability of detecting a subsequent photon in detector A[B] when it has already clicked a previous photon event is $p(A|A)$[$p(B|B)$]. The conditional probabilities $p(A|A)$ and $p(B|A)$ are given as:
\begin{align}
    p(0|0) \equiv p(A|A) = \eta_A T\Big(1 - \int_0^{\tau^A} g^{(2)}(\tau)d\tau\Big) \label{pAA}\\ 
    p(1|0) \equiv p(B|A) = \eta_B R \Big(1 - \eta_B R (\int_0^{\tau^B/2} g^{(2)}(\tau)d\tau)^2\Big) \label{pAB}
\end{align}
where $\int_0^{\tau^A} g^{(2)}(\tau)d\tau$ signifies the probability that the next incident photon is in the dead time of detector A. The second term in Eq.~\ref{pAB} signifies the probability of detector B not being in its dead time when a photon shoots into the beam-splitter, this probability is an estimation, which is based on the assumption both dead times are approximately equal. 


The probability ($p(A) \equiv p(0)$) of detector A detecting a photon is:
\begin{widetext}
\begin{align}
    p(A) &= \frac{r_A}{r_{\text{total}}} = \frac{r_A}{r_A + r_B}, \\ 
    r_A &= \eta_A T - \frac{(\eta_A T)^2 I_{\text{in}}\int^{\tau^A}_0 g^{(2)}(\tau)d\tau}{4} \\ 
    r_B &= \eta_B R - \frac{(\eta_B R)^2 I_{\text{in}}\int^{\tau^B}_0 g^{(2)}(\tau)d\tau}{4}
\end{align}
\end{widetext}

where $r_A$ and $r_B$ are the click rates of detector A and B and $I_{\text{in}}$ is the incoming intensity impinging on the beam-splitter.










 Now, we introduce $p(AB) = p(A)p(B|A)$ and $p(BA) = p(AB)$. Thus, the above equation can be modified to:
\begin{align}
    H_{\infty}(X|Y) = &-\log_2\Big(\max\{p(A) - p(AB),p(AB)\} \nonumber \\
    &+ \max\{p(AB), 1 - p(A) - p(AB)\} \Big)
\end{align}

Under the following conditions, $H_\infty(X|Y)$ yields the following results:
\begin{widetext}
    \begin{equation}
    H_\infty(X|Y) = \begin{cases}
    -\log_2(p(A)) & p(A) - p(AB) \geq p(AB) \text{ and } p(AB) \geq 1 - p(A) - p(AB),\\
    -\log_2(p(B)) & p(A) - p(AB) \leq p(AB) \text{ and } p(AB) \leq 1 - p(A) - p(AB),\\
    -\log_2(2p(AB)) & p(A) - p(AB) \leq p(AB) \text{ and } p(AB) \geq 1 - p(A) - p(AB),\\
    -\log_2(1 - 2p(A)) & p(A) - p(AB) \geq p(AB) \text{ and } p(AB) \leq 1 - p(A) - p(AB)
\end{cases}
\end{equation}
\end{widetext}

Since all four conditions have only one variable, we must have:

\begin{equation}
    H_\infty(X|Y) = -\log_2(\max\{p(A), 1-p(A), 2p(AB), 1-2p(AB)\})
\end{equation}
\appendix

 \section{Eigenbasis of the single-hole Hamiltonian}\label{first_hybr}

 Consider the $3\times3$ upper block of the matrix (\ref{electron_ion_matrix}):

\begin{align}
    V_{3\times3} = \begin{pmatrix}
         V_C & h_C & h_C \\
        h_C & V_C & h_C \\
        h_C & h_C & V_C \\
    \end{pmatrix} = (V_C-h_C)\mathrm{I} + 3h_C\mathrm{P} 
\end{align}
where $\mathrm{I}, \mathrm{P}$ are the $3\times3$ identity and projection matrices. $\mathrm{P} = 3 (\frac{1}{3} \mathbf{1}^T\mathbf{1})$, where $\mathbf{1}$ are vectors of ones. 
The eigenvalues of $(V_C-h_C)\mathrm{I}$ and $3h_C\mathrm{P}$ are $\{V_C-h_C, V_C-h_C, V_C-h_C\}$ and $\{0, 0, 3h_C\}$, respectively. 
Thus, the eigenvalues and eigenvectors of $V_{3\times3}$ are $\{V_C-h_C,V_C-h_C,V_C+2h_C\}$ and 
$\Bigg\{
\frac{1}{\sqrt{6}}\begin{pmatrix}
    2 \\ -1 \\ -1
\end{pmatrix},
\frac{1}{\sqrt{2}}\begin{pmatrix}
    0 \\ 1 \\ -1
\end{pmatrix},
\frac{1}{\sqrt{3}}\begin{pmatrix}
    1 \\ 1 \\ 1
\end{pmatrix}
\Bigg\}$, respectively. In this basis, we can rewrite the matrix in Eq.~\ref{electron_ion_matrix} as:

\begin{equation}
    V = \begin{pmatrix}
        V_C+2h_C & 0 & 0 & \sqrt{3}h_N \\
        0 & V_C-h_C & 0 & 0 \\
        0 & 0 & V_C-h_C & 0 \\
        \sqrt{3}h^*_N & 0 & 0 & V_N \\
    \end{pmatrix} 
\end{equation}
which has a block-diagonal form. Thus, we have the doubly-degenerate eigenvalue $V_C-h_C$ corresponding to the 
\gniewko{eigensubspace $\mathrm{span}\{ \frac{1}{\sqrt{6}}(2\sigma_1 - \sigma_2 - \sigma_3), \frac{1}{\sqrt{2}}(\sigma_2 - \sigma_3) \}$. } 
Now, we consider the reduced matrix:

\begin{equation}\label{2x2_matrix}
    V_{2\times2} = \begin{pmatrix}
        V_C+2h_C & \sqrt{3}h_N \\
        \sqrt{3}h^*_N & V_N
    \end{pmatrix}
\end{equation}
\gniewko{in the subspace} $\{\sigma_C(\equiv \frac{1}3(\sigma_1+\sigma_2+\sigma_3)), \sigma_N\}$. The characteristic polynomial of the matrix 
is:

\begin{align}\label{char_poly}
    \det(V_{2\times2} - \lambda\mathrm{I}_2) &= \lambda^2 
    - (V_C+2h_C+V_N) \lambda
    \nonumber \\
    &+ V_N(V_C+2h_C) - 3|h_N|^2 = 0 
\end{align}
(where $\mathrm{I}_2$ is the $2\times2$ identity matrix). On solving the characteristic polynomial given in (\ref{char_poly}), we obtain the eigenvalues ($\lambda_\pm$):

\begin{align}
    \lambda_\pm &= V_N + \frac{V_C-V_N}{2} + h_C  \pm \sqrt{\left( \frac{V_C-V_N}{2} +h_C \right)^2 + 3h_N^2} \nonumber\\
    &\equiv V_N + A \pm \Delta, 
\end{align}

where $A = \frac{V_C-V_N}{2} + h_C$ and $\Delta = \sqrt{A^2+3h_N^2}$. The corresponding orthonormal eigenvectors are found to be: 

\begin{align}
    v_- = \begin{pmatrix}
    \alpha_+ \\
    \alpha_-
\end{pmatrix} \qquad
v_+ = \begin{pmatrix}
    -\alpha_- \\
    \alpha_+
\end{pmatrix} 
\qquad
\text{where} \ \alpha_{\pm} = \sqrt{\frac{\Delta\pm A}{2\Delta}}
\label{evecs}
\end{align}

Thus, we get an eigenbasis of 4 orthonormal vectors (single electron orbitals): 
\begin{align}\label{single_electron_orbitals_app}
    \ket{x} = \frac{1}{\sqrt{6}}(2\sigma_1 - \sigma_2 - \sigma_3) \\
    \ket{y} = \frac{1}{\sqrt{2}}(\sigma_2 - \sigma_3)\label{single_electron_orbitals3} \\
    \ket{1} = \frac{\alpha_+}{\sqrt{3}}(\sigma_1+\sigma_2+\sigma_3) + \alpha_- \sigma_N \\
    \ket{2} = -\frac{\alpha_-}{\sqrt{3}}(\sigma_1+\sigma_2+\sigma_3) + \alpha_+ \sigma_N
\end{align}

\section{Spatial orbitals} \label{second_hybr} 


In this section, we show how to project tensor products of $1$-hole orbitals to obtain valid (transforming according to representations $A_1, A_2, E_1$) $2$-hole orbitals. We obtain them using the formula:

\begin{align}\label{2Hole_Wavefn}
     \Psi_r 
    &= \frac{l_r}{h} \sum_e \chi^r_e (R_e \otimes R_e) (\phi_i \otimes \phi_j) 
\end{align}
where $\phi_i, \phi_j$ are single hole orbitals [see Eqs.~\ref{single_electron_orbitals} -~\ref{single_electron_orbitals3}]. They transform under group elements as follows:


\begin{table}[htp]
\begin{tabular}{|l|l|l|}
\hline
$I$ & $\ket{x}$ & $\ket{y}$ \\ \hline
$C^+_3$ & $-\frac{1}{2}\ket{x} + \frac{\sqrt{3}}{2}\ket{y}$ & $-\frac{\sqrt{3}}{2}\ket{x} - \frac{1}{2}\ket{y}$ \\ \hline
$C^-_3$ & $-\frac{1}{2}\ket{x} - \frac{\sqrt{3}}{2}\ket{y}$ & $\frac{\sqrt{3}}{2}\ket{x} - \frac{1}{2}\ket{y}$ \\ \hline
$\sigma_v$ & $\ket{x}$  & $-\ket{y}$ \\ \hline
$\sigma_v'$ & $-\frac{1}{2}\ket{x} + \frac{\sqrt{3}}{2}\ket{y}$ & $\frac{\sqrt{3}}{2}\ket{x} + \frac{1}{2}\ket{y}$ \\ \hline
$\sigma_v''$ & $-\frac{1}{2}\ket{x} - \frac{\sqrt{3}}{2}\ket{y}$ & $-\frac{\sqrt{3}}{2}\ket{x} + \frac{1}{2}\ket{y}$ \\ \hline
\end{tabular}
\end{table}

\noindent and the orbitals $\ket{1}, \ket{2}$ are invariant under all the group operations in $C_{3v}$.
The characters $\chi^r$ for subsequent representations are given by the equations (\ref{char_A1}-\ref{char_E1}). Using the formula (\ref{2Hole_Wavefn}) we find the following projections for all tensor products of the single orbital wavefunctions $\ket{x}, \ket{y}, \ket{1}$ for subsequent representations: 
\begin{table}[htp]
\begin{tabular}{|l|l|l|l|l|l|l|}
\hline
 & $\ket{x}\otimes\ket{y}$ & $\ket{x}\otimes \ket{x}$ & $\ket{y} \otimes \ket{y}$ & $\ket{1}\otimes \ket{x}$ &  $\ket{1} \otimes \ket{y}$ & $\ket{1} \otimes \ket{1}$ \\ \hline
$A_1$ & 0 & $\frac{1}{2}(\ket{x}\otimes \ket{x} + \ket{y}\otimes \ket{y})$ & $\frac{1}{2}(\ket{x}\otimes \ket{x} + \ket{y}\otimes \ket{y})$ & 0 & 0 & $\ket{1} \otimes \ket{1}$ \\ \hline
$A_2$ & $\frac{1}{2}(\ket{x}\otimes \ket{y} - \ket{y}\otimes \ket{x})$ & 0 & 0 & 0 & 0 & 0 \\ \hline
$E_1$ & $\frac{1}{2}(\ket{x}\otimes \ket{y} + \ket{y}\otimes \ket{x})$ & $\frac{1}{2}(\ket{x}\otimes \ket{x} - \ket{y}\otimes \ket{y})$ & $-\frac{1}{2}(\ket{x}\otimes \ket{x} - \ket{y}\otimes \ket{y})$ & $\ket{1} \otimes \ket{x}$ & $\ket{1} \otimes \ket{y}$ & 0 \\ \hline
\end{tabular}
\end{table}

\noindent(results for $\ket{y}\otimes\ket{x}$, $\ket{1}\otimes\ket{x}$ and $\ket{1}\otimes\ket{y}$ has been ommited since they are easily obtainable from the results for $\ket{x}\otimes\ket{y}$, $\ket{x}\otimes\ket{2}$ and $\ket{y}\otimes\ket{1}$ by particle exchange). Removing duplicates (sometimes different product states have the same projection) we obtain 2 states in the $A_1$ representation, 1 state in $A_2$ representation and 6 states in the $E_1$ representation. The subsequent representations can be further divided into symmetric and antisymmetric sector, as summarised in the table \ref{tab:orbitals}.



\section{Spin-orbitals and their representations}
\label{spin-orbitals}

The following set of SU(2) matrices is a representation of the
$C_{3v}$ group in the spin-$\frac 12$ space:
\begin{align}
U_I = 
\begin{pmatrix}
    1 & 0 \\
    0 & 1 
\end{pmatrix},
U_{C_3^+} = 
\begin{pmatrix}
    \omega & 0 \\
    0 & \omega^* 
\end{pmatrix},
U_{C_3^-} = 
\begin{pmatrix}
    \omega^* & 0 \\
    0 & \omega 
\end{pmatrix},
U_{\sigma_v} = 
\begin{pmatrix}
    0 & 1  \\
    -1 & 0
\end{pmatrix},
U_{\sigma_v'} =
\begin{pmatrix}
    0 &\omega  \\
    -\omega^* & 0
\end{pmatrix},
U_{\sigma_v''} =
\begin{pmatrix}
    0 &\omega^*  \\
    -\omega & 0
\end{pmatrix}
\end{align}
where $\omega^{\pm i} = \exp(\pm i \frac{2\pi}{3})$.

We obtain the following actions of the group elements on the 
canonical basis of $\mathbb{C}^2 \otimes \mathbb{C^2} space$:
\begin{table}[htp]
\begin{tabular}{|c|c|c|c|c|}
\hline
    & $\ket{\uparrow\uparrow}$ & $\ket{\downarrow \downarrow}$ & $\ket{\uparrow \downarrow}$ & $\ket{\downarrow \uparrow}$ \\ \hline
    $\begin{pmatrix}
        \omega^i & 0 \\
        0 & \omega^{-i} 
    \end{pmatrix}$ & $\omega^{2i}\ket{\uparrow\uparrow}$ & $\omega^{-2i}\ket{\downarrow\downarrow}$ & $\ket{\uparrow\downarrow}$ & $\ket{\downarrow \uparrow}$ \\ 
    $\begin{pmatrix}
        0 & \omega^i \\
        -\omega^{-i} & 0 
    \end{pmatrix}$ & $\omega^{-2i}\ket{\downarrow\downarrow}$ & $\omega^{2i}\ket{\uparrow\uparrow}$ & $\ket{\downarrow\uparrow}$ & $\ket{\uparrow \downarrow}$ \\ \hline
\end{tabular}
\end{table}

Now we apply the projection formula
\begin{displaymath}
     \Psi_r = \frac{l_r}{h} \sum_e \chi^r(e) \cdot (U_e \otimes U_e) (\xi_i \otimes \xi_j)
\end{displaymath}
by summing the rows of the above table multiplied by values of element characters, appropriate for the subsequent representations:

\begin{table}[htp]
\begin{tabular}{|c|c|c|c|c|}
\hline
    & $\ket{\uparrow\uparrow}$ & $\ket{\downarrow \downarrow}$ & $\ket{\uparrow \downarrow}$ & $\ket{\downarrow \uparrow}$ \\ \hline
    $A_1$ & 0 & 0 & $\ket{\uparrow\downarrow} + \ket{\downarrow\uparrow}$ & $\ket{\downarrow\uparrow} + \ket{\uparrow\downarrow}$ \\
    $A_2$ & 0 & 0 & $\ket{\uparrow\downarrow} - \ket{\downarrow\uparrow}$ & $\ket{\downarrow\uparrow} - \ket{\uparrow\downarrow}$ \\
    $E_1$ & $\ket{\uparrow\uparrow}$ & $\ket{\downarrow\downarrow}$ & & \\
    \hline
\end{tabular}
\end{table}

The operations of the group elements are shown in the top 2 rows of the table. As an example, we show the action of the group operator to the spin state ($\uparrow\uparrow$) up to normalization:

\begin{align}
    \Psi_r(A_1) &= \ket{\uparrow\uparrow} + \omega^{2i}\ket{\uparrow\uparrow} + \omega^{-2i}\ket{\uparrow\uparrow} + \ket{\downarrow\downarrow} + \omega^{2i}\ket{\downarrow\downarrow} + \omega^{-2i}\ket{\downarrow\downarrow} \nonumber \\
    &= \ket{\uparrow\uparrow}(1 + \omega^{2i} + \omega^{-2i}) + \ket{\downarrow\downarrow}(1 + \omega^{2i} + \omega^{-2i}) = 0 \\
    \Psi_r(A_2) &= \uparrow\uparrow + \omega^{2i}\ket{\uparrow\uparrow} + \omega^{-2i}\ket{\uparrow\uparrow} - \ket{\downarrow\downarrow} - \omega^{2i}\ket{\downarrow\downarrow} - \omega^{-2i}\ket{\downarrow\downarrow} \nonumber \\
    &= \ket{\uparrow\uparrow}(1 + \omega^{2i} + \omega^{-2i}) - \ket{\downarrow\downarrow}(1 + \omega^{2i} + \omega^{-2i}) = 0 \\
    \Psi_r(E_1) &= \ket{\uparrow\uparrow}(2 - \omega^{2i} - \omega^{-2i}) = \ket{\uparrow\uparrow} \\    
\end{align}
In this way the 2-hole spin space become divided into three subspaces 
$\{\frac 1{\sqrt 2} ( \ket{\uparrow\downarrow} + \ket{\downarrow\uparrow})\}$, 
$\{\frac 1{\sqrt 2} ( \ket{\uparrow\downarrow} - \ket{\downarrow\uparrow})\}$ and 
$\{\frac 1{\sqrt 2} ( \ket{\uparrow\uparrow} \pm \ket{\downarrow\downarrow})\}$ 
transforming as $A_1$, $A_2$ and $E_1$ representation respectively.

A tensor product of two representations is in general a direct sum of other representations. We identify them performing pointwise multiplication of their characters and then decomposing the product into characters. For tensor product of $C_{3v}$ representations, the results are summarised in the table below:

\begin{table}[htp]
  \begin{tabular}{|c||c|c|c|}
\hline
$\otimes$ & $A_1$ & $A_2$ & $E_1$ \\
\hline \hline
$A_1$ & $A_1$ & $A_2$ & $E_1$ \\
\hline
$A_2$ & $A_2$ & $A_1$ & $E_1$ \\
\hline
$E_1$ & $E_1$ & $E_1$ & $A_1 \oplus A_2 \oplus E_1$ \\
\hline
\end{tabular}
\end{table}

Almost all products decompose into one-element direct sums, except $E_1 \otimes E_1$ decomposing as $A_1 \oplus A_2 \oplus E_1$ according to $(2,-1,0) \cdot (2,-1,0) = (4,1,0) = (1,1,1) + (1,1,-1) + (2,-1,0)$.

To obtain the representation structure of the whole 2-hole Hilbert space, we perform tensor product of $\mathbb{C}^3\otimes\mathbb{C}^3 = A_1^+ \oplus A_2^- \oplus E_1^+ \oplus E_1^-$ (spatial part) and $\mathbb{C}^2\otimes\mathbb{C}^2 = A_1^- \oplus A_2^+ \oplus E_1^-$ (spin part) and selecting the antisymmetrical components. We obtain:
\begin{table}[htp]
\begin{tabular}{|c||c|c|c|}
\hline
$\otimes$ & $A_1^-$ & $A_2^+$ & $E_1^+$ \\
\hline \hline
$A_1^+$ & $A_1^-$ & & \\
\hline
$A_2^-$ &  & $A_1^-$ & $E_1^-$ \\
\hline
$E_1^+$ & $E_1^-$ &  &  \\
\hline
$E_1^-$ & & $E_1^-$ & $A_1^- \oplus A_2^- \oplus E_1^-$ \\
\hline
\end{tabular}
\end{table}
To properly divide $E_1^- \otimes E_1^+ = \{ \frac 1{\sqrt{2}}(\ket{1j}-\ket{j1}), j \in \{x,y\} \} \otimes \{ \frac 1{\sqrt 2} ( \ket{\uparrow\uparrow} \pm \ket{\downarrow\downarrow}) \}$ into $A_1^- \oplus A_2^- \oplus E_1^-$ we will apply the projection formula: 


\begin{equation}\label{2hole_withspin_Wavefn}
     \Psi_r = \frac{l_r}{h} \sum_e \chi^r(e) \cdot (R_e \otimes R_e) (\phi_i \otimes \phi_j) (U_e \otimes U_e) (\xi_i \otimes \xi_j)
\end{equation}
As an example, we show the derivation of the decomposition into $A^-_1 \oplus A^-_2 \oplus E^-_1$ for the state $\frac{1}{2}(\ket{1x} - \ket{x1}) \otimes (\ket{\uparrow\uparrow} \pm \ket{\downarrow\downarrow})$. The table below shows the transformation of the state under the group operations:
\begin{center}
\begin{tabular}{|c|c|}
\hline
element & state \\
\hline \hline
$I$ & $\frac{1}{2}(\ket{1x} - \ket{x1}) \otimes (\ket{\uparrow\uparrow} \pm \ket{\downarrow\downarrow})$ \\
\hline
$C^+_3$ & $\frac{1}{2}(\ket{1}(-\frac{1}{2}\ket{x} + \frac{\sqrt{3}}{2}\ket{y}) - (-\frac{1}{2}\ket{x} + \frac{\sqrt{3}}{2}\ket{y})\ket{1}) \otimes (\omega^2\ket{\uparrow\uparrow} \pm \omega^{-2}\ket{\downarrow\downarrow})$ \\
\hline
$C^-_3$ & $\frac{1}{2}(\ket{1}(-\frac{1}{2}\ket{x} - \frac{\sqrt{3}}{2}\ket{y}) - (-\frac{1}{2}\ket{x} - \frac{\sqrt{3}}{2}\ket{y})\ket{1}) \otimes (\omega^{-2}\ket{\uparrow\uparrow} \pm \omega^{2}\ket{\downarrow\downarrow})$ \\
\hline
$\sigma_v$ & $\frac{1}{2}(\ket{1x} - \ket{x1}) \otimes (\ket{\uparrow\uparrow} + \ket{\downarrow\downarrow})$ \\
\hline
$\sigma^{'}_v$ & $\frac{1}{2}(\ket{1}(-\frac{1}{2}\ket{x} + \frac{\sqrt{3}}{2}\ket{y}) - (-\frac{1}{2}\ket{x} + \frac{\sqrt{3}}{2}\ket{y})\ket{1}) \otimes (\omega^{-2}\ket{\downarrow\downarrow} \pm \omega^{2}\ket{\uparrow\uparrow})$ \\
\hline
$\sigma^{''}_v$ & $\frac{1}{2}(\ket{1}(-\frac{1}{2}\ket{x} - \frac{\sqrt{3}}{2}\ket{y}) - (-\frac{1}{2}\ket{x} - \frac{\sqrt{3}}{2}\ket{y})\ket{1}) \otimes (\omega^{2}\ket{\downarrow\downarrow} \pm \omega^{-2}\ket{\uparrow\uparrow})$ \\
 \hline
\end{tabular}
\end{center}


Similarly, the group operations on $\frac{1}{2}(\ket{1y} - \ket{y1})\otimes(\ket{\uparrow\uparrow} \pm \ket{\downarrow\downarrow})$ can be derived. 
Following Eq.~\ref{2hole_withspin_Wavefn}, we find the decomposition of the $\frac{1}{2}(\ket{1j} - \ket{j1})\otimes(\ket{\uparrow\uparrow} \pm \ket{\downarrow\downarrow})$ into $A^-_1 \oplus A^-_2 \oplus E^-_1$ is given by:
\begin{center}
\begin{tabular}{|c|c|}
\hline
irrep & state \\
\hline \hline
$A_1$ & $\frac 12
(
(\ket{1-}-\ket{-1})\otimes\ket{\uparrow\uparrow}
-
(\ket{1+}-\ket{+1})\otimes\ket{\downarrow\downarrow}
)$ \\
\hline
$A_2$
&
$\frac 12
(
(\ket{1-}-\ket{-1})\otimes\ket{\uparrow\uparrow}
+
(\ket{1+}-\ket{+1})\otimes\ket{\downarrow\downarrow}
)$
\\ \hline
$E_1$
&
$\frac 12
(
(\ket{1+}-\ket{+1})\otimes\ket{\uparrow\uparrow}
\pm
(\ket{1-}-\ket{-1})\otimes\ket{\downarrow\downarrow}
)$\\
 \hline
\end{tabular}
\end{center}


Considering the spatial wavefunctions $\{\ket{1}, \ket{x}, \ket{y}\}$ and introducing $\ket{\pm} = \ket{x} \pm i\ket{y}$, on using Eq.~\ref{2hole_withspin_Wavefn}, we get:
\begin{center}
\begin{tabular}{|c|c|}
\hline
state & irrep
\\ \hline
\hline 
$
\left. \begin{array}{r}
\frac 1{\sqrt 2}(\ket{xx} + \ket{yy})
\\
\ket{11}
\end{array} \right\} \otimes \frac 1{\sqrt 2}
(\ket{\uparrow\downarrow} - \ket{\downarrow\uparrow})
$
&
$A_1$
\\ \hline
$
\frac 12 (\ket{xy} - \ket{yx})
\otimes
\left\{ \begin{array}{l} 
\ket{\uparrow\downarrow} + \ket{\downarrow\uparrow} \\
\ket{\uparrow\uparrow} \pm \ket{\downarrow\downarrow}
\end{array} \right.
$
&  
\begin{tabular}{@{}c@{}}
$A_1$ \\ $E_1$
\end{tabular}
\\ \hline
$
\left. \begin{array}{r} 
\ket{1j}+\ket{j1} \\
\ket{xy}+\ket{yx} \\
\ket{xx}-\ket{yy}
\end{array} \right\}
\otimes \frac 12 (\ket{\uparrow\downarrow} - \ket{\downarrow\uparrow})
$
&
$E_1$
\\ \hline
$
\frac 12 (\ket{1j}-\ket{j1}) \otimes (\ket{\uparrow\downarrow} + \ket{\downarrow\uparrow})
$
&
$E_1$
\\ \hline
$
\frac 12
(
(\ket{1+}-\ket{+1})\otimes\ket{\uparrow\uparrow}
\pm
(\ket{1-}-\ket{-1})\otimes\ket{\downarrow\downarrow}
)
$
&
$E_1$
\\
$
\frac 12
(
(\ket{1-}-\ket{-1})\otimes\ket{\uparrow\uparrow}
-
(\ket{1+}-\ket{+1})\otimes\ket{\downarrow\downarrow}
)
$
&
$A_1$
\\
$
\frac 12
(
(\ket{1-}-\ket{-1})\otimes\ket{\uparrow\uparrow}
+
(\ket{1+}-\ket{+1})\otimes\ket{\downarrow\downarrow}
)
$
&
$A_2$
\\ \hline
\end{tabular}
\end{center}
The table lists the 15 ($\{\ket{1}, \ket{x}, \ket{y}\} \otimes \{\ket{\uparrow}, \ket{\downarrow}\} \implies ^{3\times 2}C_2 \equiv 15$) total wavefunctions for a 2-hole system.

\section{Check for Background Noise}

To comment on the emission properties from $P_4$, we consider the probability of detecting an uncorrelated background noise event. The background emission from the sample's local environment, laser leakage, dark counts and electrical jitter can all be considered as background contributions leading to an increase in the value of $g^{(2)}(0)$. Since these spurious, uncorrelated events remain in the final raw bit sequence, we now understand their contribution to the conditional min-entropy $H_\infty (X|Y)$. 

We introduce the parameter $p_e$, representing the probability of
detecting a background noise event. The conditional min-entropy thus modifies to:

\begin{equation}
    H_\infty(X|Y) = -\log_2\Big(p_e + (1-p_e)f(p)\Big)
\end{equation}
where $f(p) = \max\{p(A),1-p(A),2p(AB),1-2p(AB)\}$

where $p_e = 1 - \sqrt{1-g^{(2)}(0)}$. since the fraction of single photon events is $\sqrt{1-g^{(2)}(0)}$. Here we assume that for $P_4$ $g^{(2)}(0) > 0.5$ is caused due to increased background (spurious) noise events. For $P_4$, $g^{(2)}(0) \approx
 0.8$ and $p_e = 0.2$, i.e., $H_\infty(X|Y) \to 0.1$, which is small compared to the other pillars. Therefore, if the increase in $g^{(2)}(0)$ is due to background events, randomness per quantum bit should decrease, contradictory to our experimental observations. For two emitters, all the events are from single photon sources, thereby of a source of randomness. This is why emission from $P_4$ can generate random sequences. 

 Thus, as the two-emitter system shows a higher passing proportion for the NIST subtests and the largest conditional min-entropy compared to the single emitters, we conclude that it can act as a reliable source of random numbers.